\documentclass[12pt,a4paper]{article}

\usepackage{amsmath,amssymb,graphicx,booktabs,multirow,mathtools,mathrsfs}
\usepackage[nocompress]{cite} 
\usepackage{caption,subcaption,float}
\usepackage{jheppub}
\usepackage[toc,page]{appendix}
\usepackage{color}
\usepackage{mathrsfs}

\makeatletter
\def\@fpheader{\vspace{0pt}}
\makeatother

\newcommand{\be}{\begin{equation}}
\newcommand{\ee}{\end{equation}}
\newcommand{\bal}{\begin{aligned}}
\newcommand{\eal}{\end{aligned}}

\newcommand*\dAlembert{\mathop{}\!\mathbin\Box}

\newcommand{\bes}{\begin{split}}
\newcommand{\ees}{\end{split}}
\def\bea{\begin{equation}\begin{aligned}}
\def\eea{\end{aligned}\end{equation}}

\newcommand{\veca}{\mathbf{a}}
\newcommand{\vecb}{\mathbf{b}}

\title{The Sorkin-Johnston State in a Patch of the Trousers Spacetime}

\author{Michel Buck}
 \author[a,b,c]{\!, Fay Dowker}
 \author[a]{\!, Ian Jubb\,}
 \author[c]{\!, Rafael Sorkin}
\affiliation[a]{Blackett Laboratory, Imperial College, London, SW7 2AZ, U.K.}
\affiliation[b]{Institute for Quantum Computing, University of Waterloo, ON, N2L 3G1, Canada}
\affiliation[c]{Perimeter Institute, 31 Caroline Street North, Waterloo ON, N2L 2Y5, Canada}
\abstract
{
A quantum scalar field in a patch of a fixed, topology-changing, 
$1+1$ dimensional ``trousers'' spacetime is studied using the Sorkin-Johnston formalism. The isometry group of the patch is the dihedral group, the symmetry group of the square. 
The theory is shown to be pathological in a way that can be interpreted as the topology change giving rise to a divergent energy, in agreement with previous results. In contrast to previous results,  it is shown that the infinite energy is localised not only on the future light cone of the topology changing singularity, but also on the  past cone, due to the time reversal symmetry of the Sorkin-Johnston state.
}
\notoc
\begin{document}
\maketitle
\flushbottom

\section{Introduction}\label{sec:intro}

There are good reasons to believe that topology change will play a role in
quantum gravity.  From the point of view of a gravitational sum-over-histories, dimensional analysis suggests that structures on Planckian scales will have a gravitational action of order $\hbar$, which would lead to very little suppression in the path-integral~\cite{Sorkin:1997gi}. Such considerations suggest that Planck scale topology-change, at least, should be taken into account in a quantum theory of gravity. Going further, Sorkin has argued that without topology change quantum gravity would be inconsistent, with the strongest evidence 
coming from the theory of topological geons~\cite{Sorkin:1986geons}, particles 
built on non-trivial  spatial topology. Geons suffer from violations of the spin-statistics correlation and other problems in a framework with frozen spatial topology. Allowing topology change might solve these problems and, conversely, considering how to
make the physics of geons consistent might give clues about the 
rules that govern topology change in quantum gravity~\cite{Sorkin:1996yt,Sorkin:1989ea,Dowker:1996ei}. 

At a formal level, it is easy enough to 
conceive of including topology changing manifolds in the gravitational path integral. However, a theorem of Geroch~\cite{Geroch:1967fs} tells us that a Lorentzian metric on a manifold in which the spatial topology changes must contain closed timelike curves. If one wants to avoid the 
pathologies that go along with closed timelike curves~\cite{Sorkin:1997gi,thorne1993closed}, one can consider the alternatives of metrics that are Lorentzian \emph{almost everywhere} (degenerating at a finite set of isolated points) and which retain a well-defined causal order~\cite{Sorkin:1989ea}, or, going further,  metrics with signature change~\cite{Dray:1991zz} or Euclidean signature~\cite{Gibbons:2011dh}. One can then investigate the action of a topology changing spacetime in a background field approximation by studying linear-order quantum fluctuations, or as a first step 
by investigating a free massless scalar quantum field 
in the background spacetime, a study within the framework of quantum field theory in curved spacetime.

Choosing the histories in the path integral to be Lorentzian spacetimes with well-defined causal order and isolated singularities, one is then faced with the challenge that such topology changing spacetimes are not globally hyperbolic in the usual sense. Since global hyperbolicity is a basic assumption in text book quantum field theory, this means that one is necessarily charting new territory in investigating quantum field theory 
in such spacetimes. New rules must be created and analysed to see if they are 
self-consistent and physically plausible. 

Work along these lines was carried out by Anderson and DeWitt~\cite{Anderson:1986ww}, who studied the quantum theory of a free massless scalar field on the topology-changing two-dimensional ``trousers'' spacetime, in which a circle splits into two (or vice-versa), see Figure~\ref{fig:trousers-flattened-raw}. 
This spacetime admits an almost everywhere Lorentzian metric, which is flat  everywhere except at an isolated singular point, the ``crotch singularity''.  Expanding the scalar field in terms of modes on a spacelike hypersurface in the ``in''-region and specifying a particular ``shadow rule'' to propagate the modes past the topology-changing hypersurface into the ``out''-region, Anderson and DeWitt concluded that the expectation value of the stress-energy tensor evaluated in the in-vacuum has incurable (squared Dirac-delta) divergences on the light-cone of the singularity. They argued that this means that the trousers-type topology-change is dynamically forbidden. Manogue et al.~\cite{Copeland:1988tr} revisited the problem with a more careful analysis. They argued that the propagation rule of Anderson and DeWitt is unphysical because the Klein-Gordon product  is not conserved when using the shadow rule to propagate solutions past the topology-changing hypersurface. Deriving a one-parameter family of propagation laws that conserve the inner product they arrived, nevertheless, at the same conclusion: an infinite burst of energy emanating from the singularity.



Recently a new approach to QFT has been proposed by Sorkin~\cite{Sorkin:2011pn,Afshordi:2012jf}
based on work by Johnston on QFT on a causal set~\cite{Johnston:2009fr}.
In this paper we apply the  Sorkin-Johnston (SJ) formalism to the trousers, not only to see what light it might shed on previous 
results, but also as an exercise in the  new approach. The starting point of the SJ approach for a free scalar field
is the retarded Green function, rather than the field operator as a solution of
the equations of motion. The Green function leads to
a distinguished quantum state --- a candidate ``ground state'' ---  for 
a spacetime region without further input. In a globally hyperbolic spacetime the retarded Green function is unique but in a topology changing spacetime we expect that there will be a choice of Green functions. This turns out to be the case and we will see that there 
is a separate QFT for each choice. 





\section{Background and Setup}\label{Backeground and Setup}

\subsection{The SJ Formalism}\label{sec:the_sj_formalism}

Here we give a brief review of the SJ formalism~\cite{Sorkin:2011pn,Afshordi:2012jf}
for a free scalar field, $\phi$, in a globally hyperbolic spacetime, $(M,g_{\mu\nu})$, of finite volume. Given the retarded Green function, $G(x,y)$, the Pauli-Jordan function is defined as $\Delta(x,y)=G(x,y)-G(y,x)$ ($x$ and $y$ are spacetime points). Note that $\Delta(x,y)$ is antisymmetric. In a globally hyperbolic spacetime, the transpose of 
the retarded Green function is the advanced Green function and so 
$\Delta(x,y)$ is a solution of the equations of motion in both its arguments. 
We will see that this condition will need to be imposed by hand in the 
trousers spacetime, as the connection between retarded and advanced 
Green functions is not automatic. 

The Hilbert space ${L}^2(M)$ of equivalence classes of complex functions on $(M,g_{\mu\nu})$  has inner product
\be
\langle [f],[g]\rangle:=\int_{M}dV_x f(x)^*g(x)
\ee
where $[f], [g] \in{L}^2(M)$ (square brackets denote equivalence classes and $*$ denotes complex conjugation). In what follows we will abuse notation and refer to an element of the Hilbert space by one of its representative functions.

We define the \emph{Pauli-Jordan operator} as an operator on 
the Hilbert space which is given by the integral operator on representative 
functions whose kernel is the Pauli-Jordan function $\Delta(x,y)$:
\be
(\Delta f)(x)=\int_ M dV_y \Delta(x,y)f(y).
\label{eq:ideltadef}
\ee
Assuming that $\Delta(x,y)$ is a square integrable kernel, i.e. that $\Delta(x,y)\in {L}^2(M\times M)$, then the operator $i\Delta$ is a self-adjoint Hilbert-Schmidt operator~\cite[Thm. VI.23]{RS} and the spectral theorem for such operators says that $i\Delta$ has a set of real eigenvalues $\lambda_\veca$ and a complete orthonormal set of eigenfunctions $\mathfrak u_\veca$ which satisfy
\be
i\Delta \mathfrak u_\veca=\lambda_\veca \mathfrak u_\veca,
\qquad
\lambda_\veca\in\mathbb R.
\label{eq:eigenveceq}
\ee
Since $\Delta(x,y)$ is a real function, it follows that
\be
i\Delta \mathfrak u_\veca=\lambda_\veca \mathfrak u_\veca \implies i\Delta \mathfrak u^*_\veca=-\lambda_\veca \mathfrak u^*_\veca,
\ee
which means that for the non-zero eigenvalues, the eigenfunctions of $i\Delta$ come in pairs:
\be
i\Delta \mathfrak u^\pm_\veca=\pm\lambda_\veca \mathfrak u^\pm_\veca,
\ee
where $\lambda_\veca>0$ and $\mathfrak u^-_\veca=\mathfrak u^{+*}_\veca$. Moreover, these eigenfunctions (appropriately normalised) are orthonormal in the ${L}^2(M)$ inner product:
\bea
\langle \mathfrak u^\pm_{\veca\vphantom{\vecb}},\mathfrak u^\pm_{\vecb}\rangle&=\delta_{\veca\vecb}\\
\langle \mathfrak u^+_{\veca\vphantom{\vecb}},\mathfrak u^-_{\vecb}\rangle&=0.
\label{eq:SJorthonormality}
\eea
$i\Delta(x,y)$ is the sum of its positive and negative parts:
\be\label{idelta_q_decomp}
i\Delta(x,y )= Q(x,y )-Q(x,y )^*,
\ee
where
\be
Q(x,y )=\sum_\veca\lambda^{}_\veca \mathfrak u_\veca^+(x)\mathfrak u_\veca^-(y ).
\label{eq:Q}
\ee
The SJ state is the pure Gaussian state defined by its Wightman function,
\be
W_{SJ}(x,y):=Q(x,y)=\sum_\veca\lambda^{}_\veca \mathfrak u_\veca^+(x)\mathfrak u_\veca^-(y )\,.
\label{eq:W}
\ee


Although the topology changing spacetime we will look at 
is not globally hyperbolic in the usual sense, it does have a well-defined causal structure so that the 
notion of retardedness of a Green function makes sense, and 
it has finite volume so the SJ formalism can 
be extended to our case if an appropriate Green function can be found.

\subsection{The Trousers Spacetime}\label{sec:qftintrousers}

\begin{figure}[t!]
\centering
\includegraphics[
clip=true,
width=0.44\textwidth]
{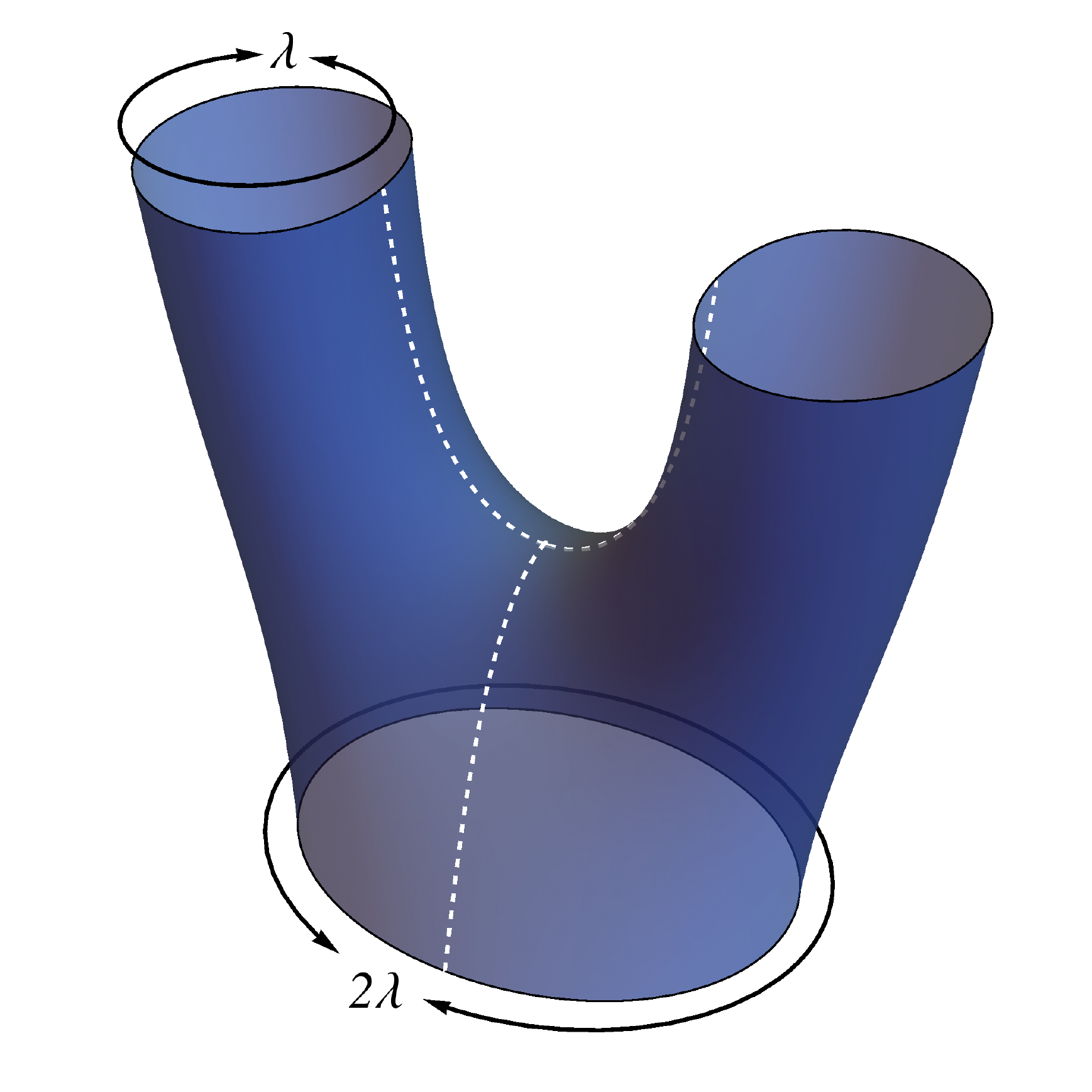}
\hspace{0.1\textwidth}
\includegraphics[
clip=true,
width=0.45\textwidth]
{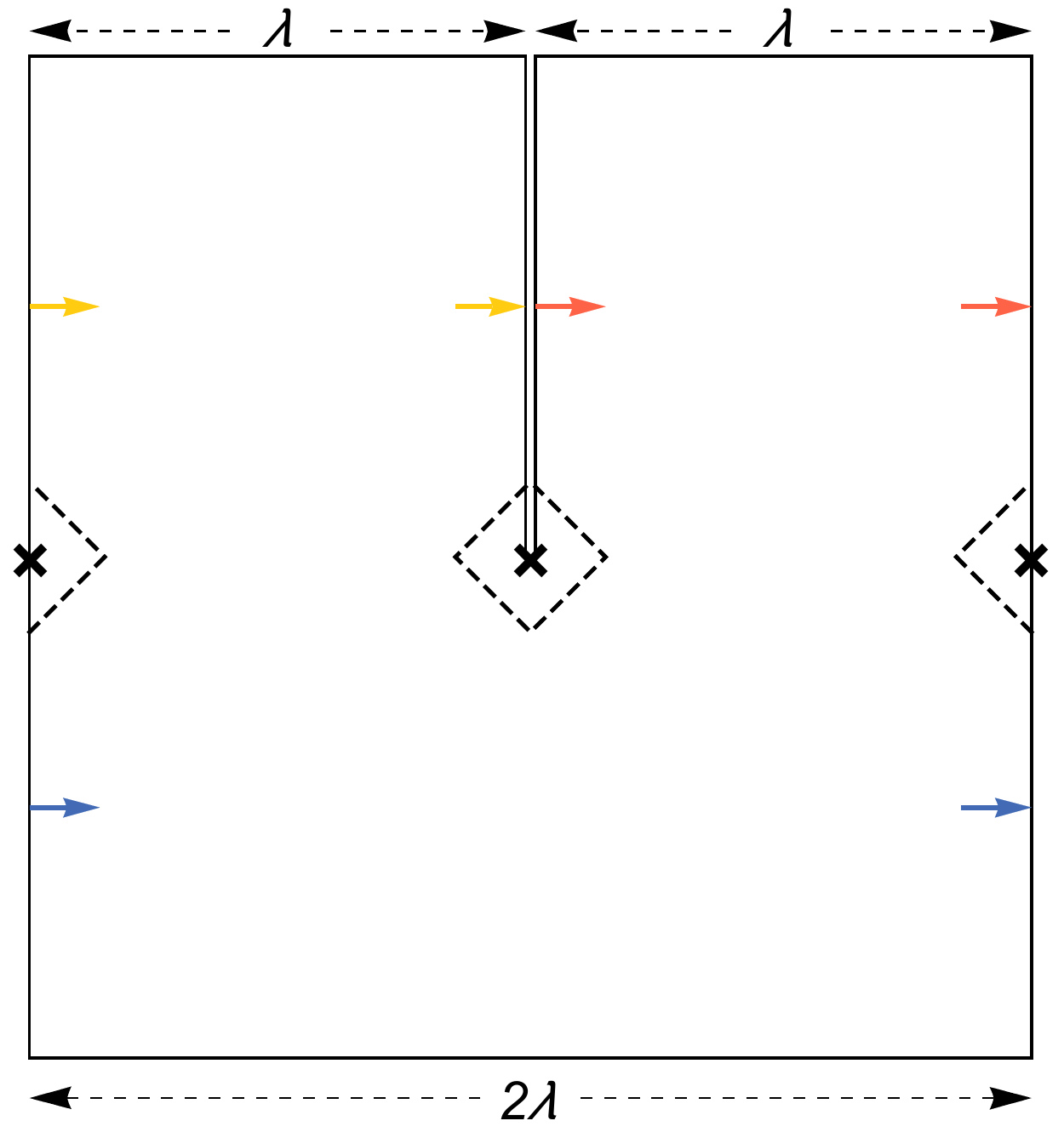}
\caption{The trousers spacetime is shown on the left. The flat two-dimensional representation of the trousers on the right is obtained by cutting along the dotted lines and unwrapping the trousers on the left. The arrows indicate the respective identifications in the trunk and in the left and right legs. The crosses are identified and  mark the location of $x_c$, the singularity. The dashed lines on the right form the boundary of a neighbourhood of $x_c$ which we call the pair of diamonds.}
\label{fig:trousers-flattened-raw}
\end{figure}

Keeping with tradition, let us hang the trousers upside down as in Figure \ref{fig:trousers-flattened-raw} and use Cartesian coordinates $(T,X)$ in which $T=0$ separates the ``legs'' and the ``trunk''. The spatial coordinate $X$ lies in the range $[-\lambda,\lambda]$ and the singularity, $x_c$ lies at the origin: $x_c=(0,0)$. The coordinates in the trunk extend to coordinates in the left and right legs, i.e. we identify points $(0+, X)$ in the legs with points $(0-,X)$ in the trunk for $X\neq0$.  In the trunk, i.e. for $T<0$, we identify $X=-\lambda$ with $X=\lambda$. In the legs $T>0$. 
 In the left leg we identify $X=-\lambda$ with $X=0-$ and in the right leg, we identify $X=\lambda$ with $X=0+$. The metric on the trousers is locally flat everywhere except at $x_c$ where it is degenerate.\\
 
To build the SJ state in the trousers we need to identify the positive eigenvalue eigenfunctions of $i\Delta$ as in the analysis of the flat causal diamond~\cite{Afshordi:2012ez}. For this, we need the Pauli-Jordan function $\Delta(x,y)=G(x,y)-G(y,x)$, and thus the retarded Green function in the trousers.

One way in which Green functions in the trousers differ from those in Minkowski space
is due to the cylindrical topology 
of the trunk and legs. Consider the quantum field theory on a flat cylinder $S^1\times\,\mathbb R$ (no topology-change). The future and past light-cones of any point, $x$, will wrap around the cylinder and intersect at a set of conjugate points. This means that the retarded Green function on the cylinder is not equal to the
retarded Green function $G_{\mathsf{Mink}}(x,y)$ of two-dimensional Minkowski space.  At the first conjugate point to the past of $x$, call it $x'$, there is a contribution  $-\delta^{(2)}(x-x')$ to
$\dAlembert_x G_{\mathsf{Mink}}(x,y)$. The Green function on the cylinder is obtained by adding to $G_{\mathsf{Mink}}(x,y)$ appropriate multiples of $G_{\mathsf{Mink}}(x',y)$ for every conjugate point $x'$: the usual method of images.

\begin{figure}[t!]
\centering
\includegraphics[
clip=true,
width=0.9\textwidth]
{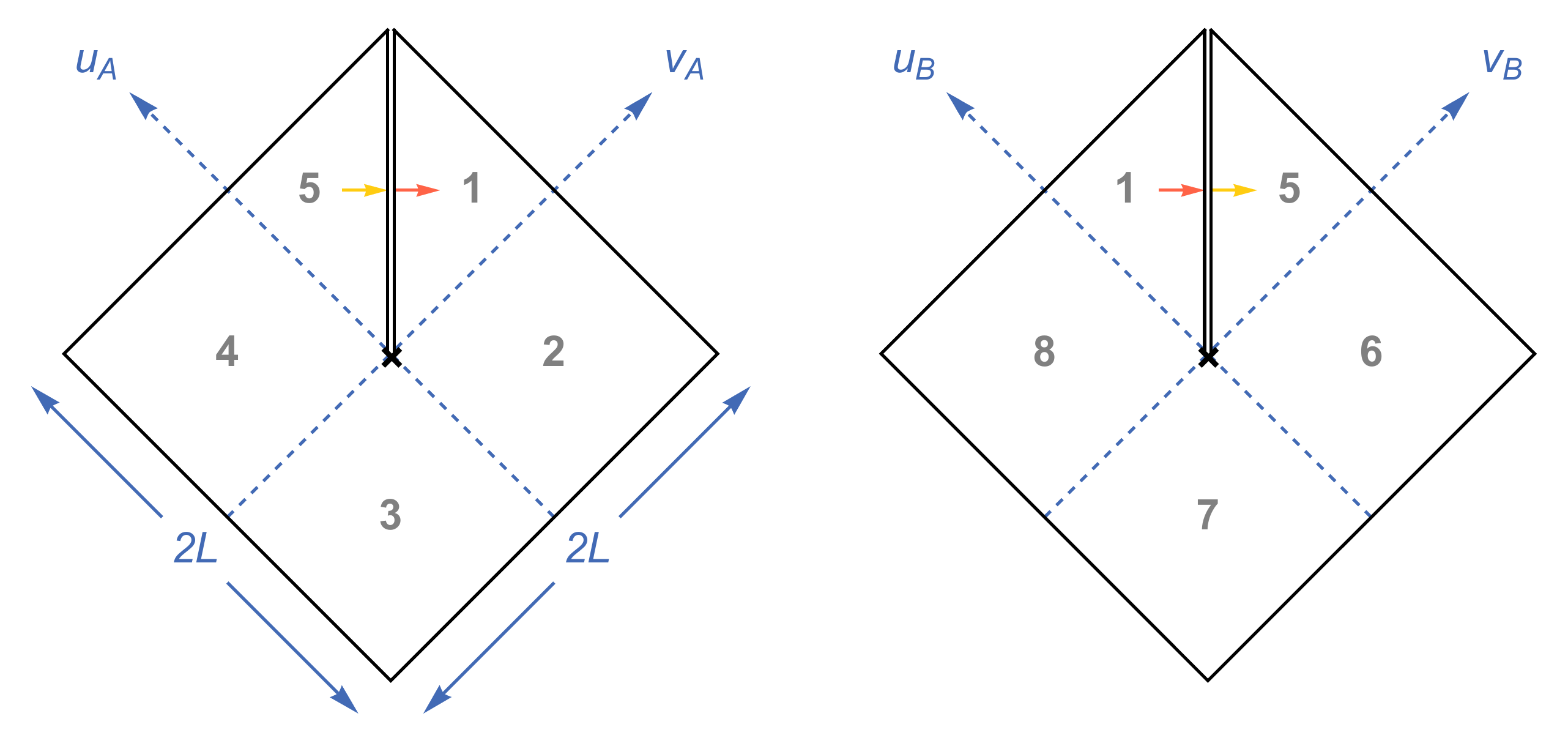}
\caption{The pair of diamonds in more detail. Diamond $A$ is on the left and 
diamond $B$ is on the right. The  arrows  in regions 1 and 5 indicate the topological identifications inherited from the trousers. The dashed lines are the past and future lightcones from the singularity.}
\label{fig:pairofdiamonds}
\end{figure}
\begin{figure}[t!]
\centering
\includegraphics[
clip=true,
width=0.45\textwidth]
{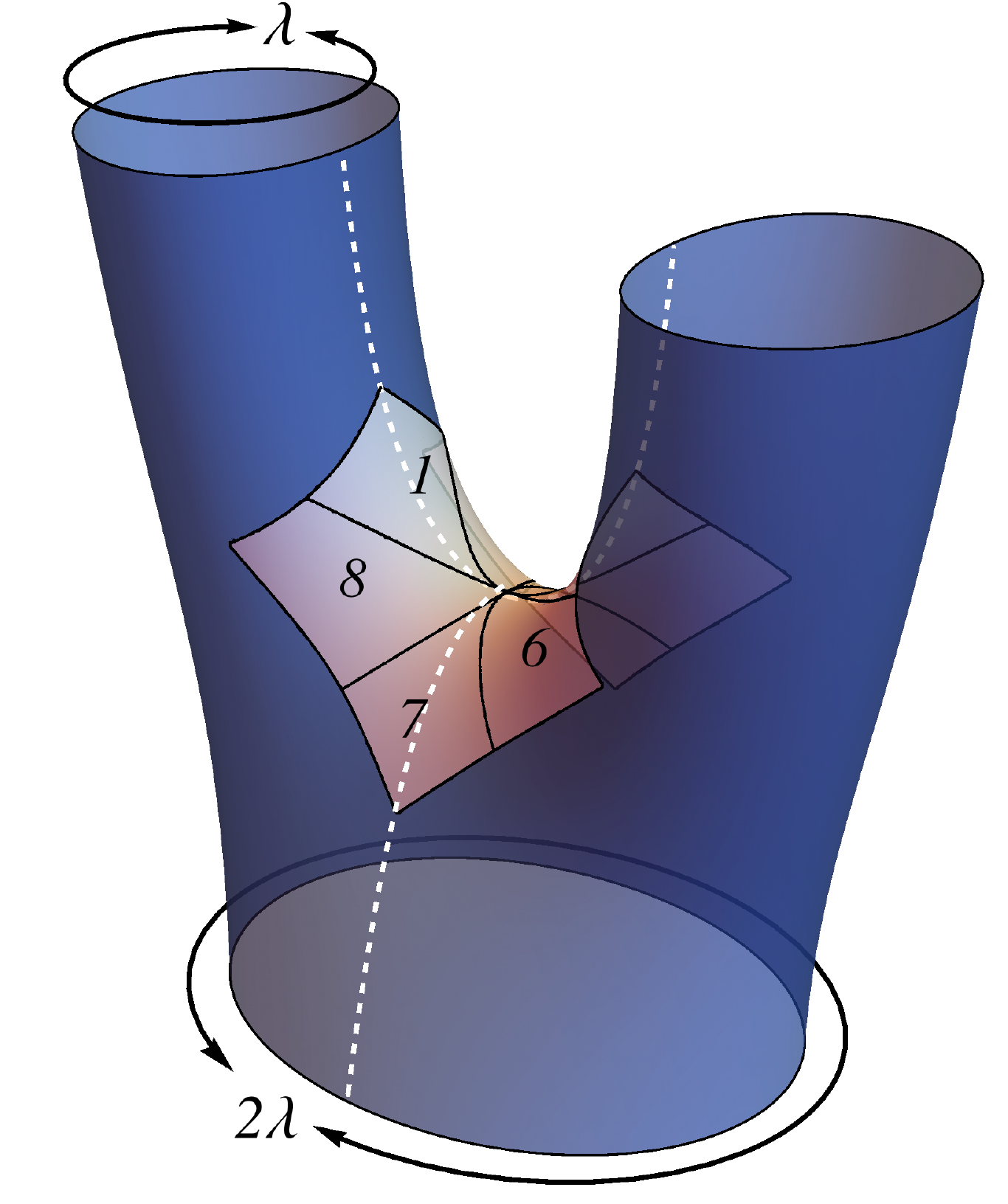}
\caption{The pair of diamonds on the trousers. The numbers illustrate the different regions of the pair of diamonds.}
\label{fig:trousers-with-diamond}
\end{figure}

In order to isolate the features of the trousers spacetime that are most pertinent to the physics of topology-change, we could restrict ourselves to a thin enough slab of the trousers containing the singularity such that no wrapping around occurs, e.g. $|T|\leq T_{max}$ for some $T_{max}<\frac\lambda4$. However, it will be most convenient to restrict further to a smaller neighbourhood of the singularity. Consider, therefore, two points, one in the left and one in the right leg, each lying directly above the singularity: $x_{leg}^\pm=(T_0, 0\pm)$. Consider the intersection of the union of their causal pasts with the causal future of two points in the trunk, $x^+_{trunk}=(-T_0,0)$ and $x^-_{trunk}= (T_0, \lambda)$, each of which lies directly below the singularity. This region consists of the two diamonds outlined with dashed lines in Figure~\ref{fig:trousers-flattened-raw}. We refer to this spacetime as the \emph{pair of diamonds}. Figure~\ref{fig:pairofdiamonds} shows the pair of diamonds, with the topological identifications inherited from the trousers.  When the two diamonds are depicted next to each other as in Figure~\ref{fig:pairofdiamonds}, the left diamond ($A$) corresponds to the diamond seen in the centre of the cut open trousers (the right diagram in Figure \ref{fig:trousers-flattened-raw}) and the right diamond ($B$) is made up of the two halves at the sides of the cut open trousers. Figure~\ref{fig:trousers-with-diamond} shows how the pair of diamonds embeds in the original picture of the trousers. The pair of diamonds spacetime captures the essential causal structure of the trousers topology change. 

\subsection{The Pair of Diamonds}\label{The Pair of Diamonds}


In order to discuss the pair of diamonds, $\cal{M}$, and functions on it, 
it will be useful to have a coordinate system that respects the symmetry between the 
two diamonds, $A$ and $B$. 
We will use both Cartesian, $(T_i,X_i)$, and light-cone coordinates, $(u_i,v_i)$ (where $u_i=\frac{1}{\sqrt{2}}(T_i-X_i)$ and $v_i=\frac{1}{\sqrt{2}}(T_i+X_i)$),  and subscripts $i = A,B$,  refer to the corresponding 
diamond. The trousers coordinates (without subscript) defined previously and the coordinates on the two diamonds are related as follows. The coordinate system on diamond $A$ agrees with the trousers coordinate system since they have the same origin: $T_A=T,\, X_A=X\,, u_A=u, v_A=v$. On diamond $B$, the left side comes from the right edge of the trousers and the right side comes from the left edge of the trousers, so the relations between the coordinate systems are 
\be
\begin{aligned}
T_B &= T\\
X_B&=X-\lambda\quad\text{for}\;X>0\\
X_B&=X+\lambda\quad\text{for}\;X<0
\end{aligned}
\quad\iff\quad
\begin{aligned}
\left.
\begin{aligned}
u_B&=u+\lambda/\sqrt2\\
v_B&=v-\lambda/\sqrt2
\end{aligned}
\quad\right\}
\quad\text{for}\;X>0\hphantom{.}\\
\left.
\begin{aligned}
u_B&=u-\lambda/\sqrt2\\
v_B&=v+\lambda/\sqrt2
\end{aligned}
\quad\right\}
\quad\text{for}\;X<0.
\end{aligned}
\ee
The coordinate range for the light-cone coordinates on each diamond is $[-L,L]$ where $\sqrt{2} L<\lambda/2$. In both the $A$ and $B$ coordinate systems the singularity, 
$x_c$,  is at the origin of coordinates. For $0<T_A,T_B<\sqrt{2}L$, we identify $X_A=0^-$ with $X_B=0^+$ and vice versa. The two coordinate systems do not correspond to a split into left and right legs in the trousers manifold: for example, both the top left part of diamond $A$ (i.e. $u_A>v_A>0$) and the top right part of diamond 2 (i.e. $v_B>u_B>0$) belong to the left leg of the trousers.

We will use notation $x$, $y$ without subscripts to denote general 
points in the  manifold and use  indicator functions to restrict support 
of functions onto subregions. We define $\chi_R(x)$ to be the function that is $1$ when $x\in R$ and zero otherwise. We define eight regions, $R_i$, where $i=1,...,8$, 
whose boundaries are the 
past and future null lines from the singularity, as shown in Figure \ref{fig:pairofdiamonds}.  For definiteness we choose the regions to include their
boundaries so that their union is the whole manifold minus the singularity $x_c$, 
but we could choose them to be open or assign the boundary points to exactly one of the regions. This does not make a difference, as we are working in $L^2(\mathcal{M})$. 

For convenience we write the corresponding indicator functions as $\chi_i(x) := \chi_{R_i}(x)$. We will also use notation $\chi_{1,2}(x) := \chi_1(x) + \chi_2(x)$ and $\chi_{2,3,5}(x) := \chi_2(x) + \chi_3(x)+ \chi_5(x)$
\textit{etc.} to denote the indicator functions for unions of these regions. 

We consider the singularity as a point of spacetime. The
 metric degenerates at the singularity but the pair of diamonds
spacetime including the singularity nevertheless possesses a 
natural, well defined causal order. For example the singularity $x_c$ 
is to the causal past (future) of all points in and on the boundaries of regions 
1 and 5 (3 and 7) in Figure \ref{fig:pairofdiamonds}. We denote
the causal order by $\preceq$ where $y \preceq x$ (equivalently,
$x \succeq y$) means that $y$ is
in the causal past of $x$. We denote by $[x,y]$ the causal interval, 
$[x,y] = \{ z \in {\cal{M}}\,|\, x\succeq z \succeq y \}$. 

\subsection{Isometries of the Pair of Diamonds}\label{Isometries of the Pair of Diamonds}

The isometry group for the pair of diamonds is generated by two transformations, one of which can be thought of as a ``parity'' transformation and the other as a ``time reversal''.
The parity transformation, $\mathfrak{P}: \mathcal{M}\rightarrow\mathcal{M}$, is the isometry that reflects both diamonds, $A$ and $B$, each in its own vertical axis
of symmetry. To define the time reversal map, $\mathfrak{T}:\mathcal{M}\rightarrow\mathcal{M}$, we need only specify its action on a single region $R_i$ and that fixes its action on the other regions by continuity. We choose to specify the action of $\mathfrak{T}$ on $R_1$ to be a reflection of $R_1$ in its own horizontal axis of symmetry followed by a translation (in the obvious sense) of $R_1$ onto $R_3$. Then the action of $\mathfrak{T}$ on the other regions is:  reflect $R_2$  in its horizontal axis;  reflect $R_3$ in its horizontal axis and translate onto $R_1$; reflect $R_4$  in its horizontal axis and translate onto $R_8$; reflect
$R_5$ in its horizontal axis and translate onto $R_7$; reflect $R_6$  in its horizontal axis; reflect $R_7$  in its horizontal axis  translate onto $R_5$; reflect $R_8$  in its horizontal axis and translate onto $R_4$.

There are actually two isometries that have an equal claim to being called ``time reversal'' on $\mathcal{M}$ and we chose one of them above to be $\mathfrak{T}$. The isometry 
that  time-reverses $R_1$ and then translates it onto $R_7$ --- instead of $R_3$ --- 
is equal to $\mathfrak{P}\circ\mathfrak{T}\circ\mathfrak{P}$.
$\mathfrak{P}$ and $\mathfrak{T}$ generate the isometry group. For example, 
the ``swap" isometry that interchanges the two diamonds, $A\leftrightarrow B$, is equal to $(\mathfrak{P}\circ\mathfrak{T})^2$.

\begin{figure}[t!]
\centering
\includegraphics[
clip=true,
width=0.5\textwidth]
{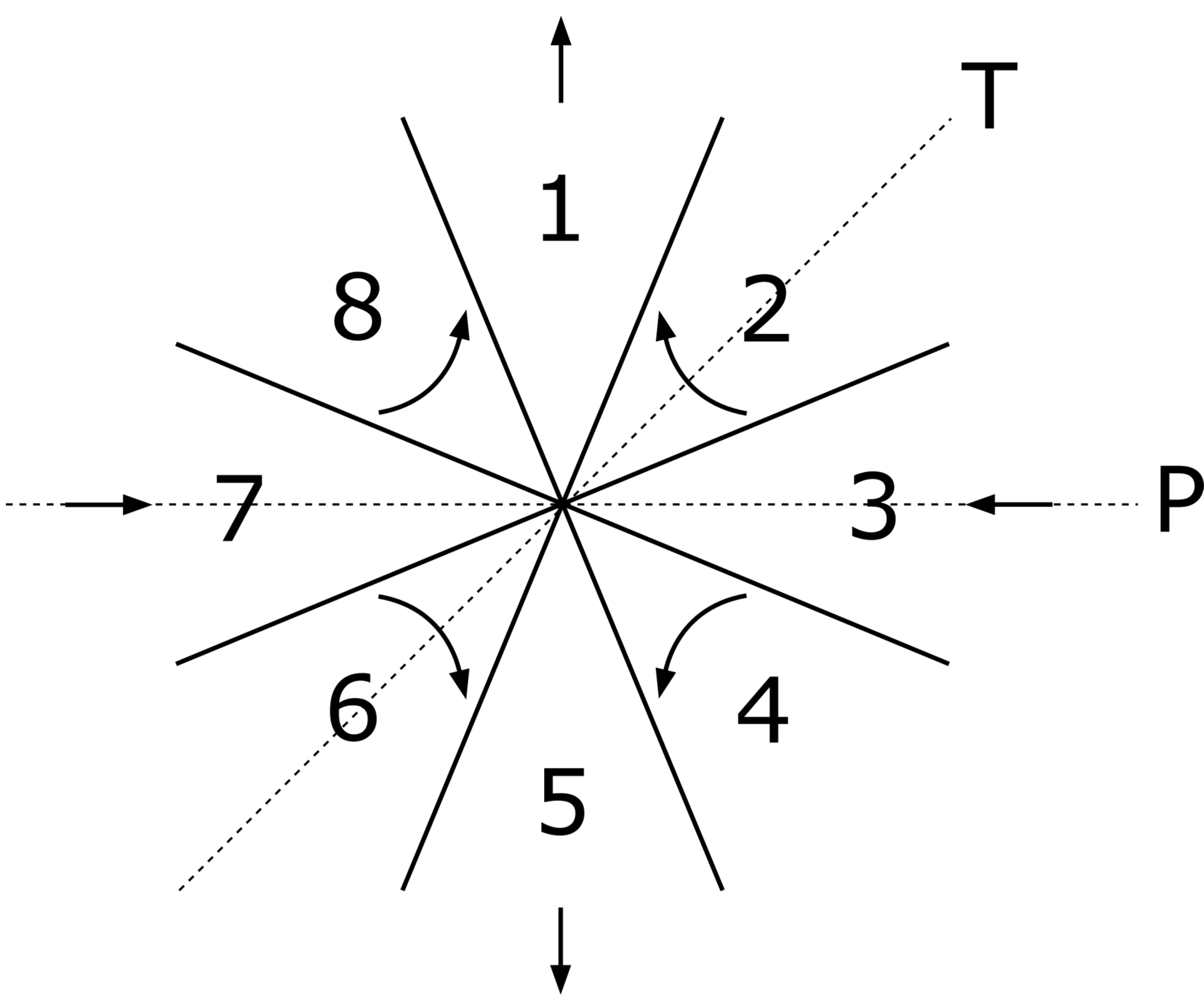}
\caption{The pair of diamonds on the trousers, as ``viewed from above''. The numbers correspond to the same 8 regions as before. The arrows represent the direction of time in each region. Isometry $\mathfrak{P}$ is reflection in the dotted horizontal line, labelled $\mathsf{P}$. Isometry $\mathfrak{T}$ is reflection about the dotted line at $45^{\circ}$ to the horizontal,   labelled $\mathsf{T}$.}
\label{fig:pod_from_above}
\end{figure}

The isometry group is the dihedral group, $D_4$, the symmetry group of the square which can be seen by viewing the trousers  in Figure~\ref{fig:trousers-flattened-raw} from above. From this point of view, the regions $R_1$ to $R_8$ are arranged as in Figure~\ref{fig:pod_from_above}. Representing topology change in this way is useful in studying the causality properties of topology change ~\cite{Dowker:1997hj}. One can determine how the parity and time reversal operations act on this representation of the spacetime, 
Figure~\ref{fig:pod_from_above}. $\mathfrak{P}$ is reflection in the horizontal dotted line marked P and $\mathfrak{T}$ is reflection in the dotted  line marked T at $45^{\circ}$ to the horizontal. 
The group $D_4$ is the symmetry group of a square and is generated by a reflection in the horizontal axis and a reflection in a diagonal. Thus, the isometry group of the pair of diamond is $D_4$.

\section{Green Functions}

\subsection{$1+1$ dimensional Minkowski}\label{$1+1$ dimensional Minkowski}

To construct the SJ theory of 
a massless scalar field, $\phi$, on the pair of diamonds, $\cal{M}$,
we must decide what it means to be a solution of the wave equation at the singularity, as the differential equation is not defined there. So let us first look at different ways to express 
the wave equation in 1+1 dimensional Minkowski space. 

The wave equation is 
\begin{equation}
\dAlembert f = 0 \label{waveeq}
\end{equation} 
so that 
\begin{equation}
\int_A dV \dAlembert f = 0 
\label{waveeqint}
\end{equation} 
for every measureable region $A$. By Stokes' theorem we have
\be
\int_A dV \dAlembert f =  \oint_{\partial A}\,d\Sigma^{\mu}\,\frac{\partial}{\partial x^{\mu}}f \,,
\ee
where the boundary $\partial A$ is traversed anti-clockwise  and  $d\Sigma_x^{\mu}$ is the normal surface element. We have implicitly assumed here that $A$ is such that its 
boundary is nice enough --- say, connected, 
non-self intersecting and piecewise smooth, for definiteness --- for this to be meaningful. If we define 
\be
\mathfrak B^{A} f := \oint_{\partial A}\,d\Sigma^{\mu}\,\frac{\partial}{\partial x^{\mu}}f 
\ee
then a solution satisfies 
$\mathfrak B^{A} f =0$ for all nice enough $A$. 

When the region is a causal interval, or \textit{causal diamond}, $D$,  this 
boundary integral only picks up the values of the function at the corners
of the diamond, because the normal derivatives in the integrand 
 become tangential when the boundary is null. The full boundary integral is a sum of the integrals along the four null segments, and each one of the integrands is a total derivative with respect to the null coordinate $u$ or $v$, so that
 \be
\int_D dV \dAlembert f=\oint_{\partial D}d\Sigma_x^{\mu}\frac{\partial}{\partial x^{\mu}}f 
= - 2 \left[f(x_1) - f(x_2) + f(x_3) - f(x_4)\right] \,,
\ee
where $x_1$ is the future tip of the diamond and the other corners are 
labelled in clockwise order. 
The boundary integral condition can therefore be written
 \be \label{ceef}
 \mathfrak C^D f= 0\,,
 \ee
for each causal diamond, $D$, where we have defined
\be 
 \mathfrak C^D f: = f(x_1) - f(x_2) + f(x_3) - f(x_4) \,.
 \ee

If $f$ is differentiable then the condition (\ref{ceef}) for all causal diamonds implies
$ \dAlembert f = 0 $ since 
\begin{align*}
\dAlembert f (u,v) &= -2 \frac{\partial}{\partial u}\frac{\partial}{\partial v}f(u,v)\\
 & = -2 \lim_{\delta u, \delta v \to 0} \frac{
 f(u+\delta u, v + \delta v) - f(u, v+ \delta v) + f(u,v) - f(u + \delta u, v) }{\delta u \delta v}\\
 & = 0 \,.
 \end{align*}

Green's equation is 
\begin{equation}
\dAlembert_x G(x,y)= \delta(x,y) \label{greenseq}
\end{equation} 
for all $x,y$, where  $\dAlembert_x$ denotes the d'Alembertian with respect to argument $x$. This means that
\begin{equation}
\int_A dV_x \dAlembert_x G(x,y) = \chi_A(y)
\label{greenseqint}
\end{equation} 
for any measureable region $A$. 

Again, Stokes' theorem gives the 
boundary integral form of the condition,
\be
\label{eq:alternativegreen}
\mathfrak B_x^A G(x,y)= \chi_A(y)
\ee
for each point $y$ and each nice enough region $A$,
where 
\be 
\mathfrak B^A_x G(x,y) : = \oint_{\partial A}d\Sigma_x^{\mu}\frac{\partial}{\partial x^{\mu}}G(x,y)\,.
\ee
 And, when the region is a causal diamond, $D$, with 
corners $x_1,\dots x_4$ as before we have 
 \be 
 \mathfrak C_x^D G(x,y)= -\frac{1}{2} \chi_D(y)\,,
 \ee
where 
\be \label{cee}
 \mathfrak C_x^D G(x,y) : = G(x_1,y) - G(x_2,y) + G(x_3,y) - G(x_4,y)\,,
 \ee
and the subscript $x$ denotes that ${\mathfrak{C}}^D_x$ 
acts on the argument $x$ of $G(x,y)$.

Similarly to the solution, the condition (\ref{cee}) for all causal diamonds and all 
points $y$ is equivalent to Green's equation. 

Finally, we note that the explicit form of the 1+1 dimensional Minkowski space retarded Green function is
 \be
 G_{\mathsf{Mink}}(x,y) = - \frac12\chi_{\succ}(x,y)\,,
 \ee
  where $\chi_{\succ}(x,y) = 1$ when $x\succ y$ and $= 0$ otherwise.

\subsection{The Pair of Diamonds}

Consider now the massless scalar field theory on the pair of diamonds, $\cal{M}$. 



\begin{figure}[t!]
\centering
\includegraphics[
clip=true,
width=\textwidth]
{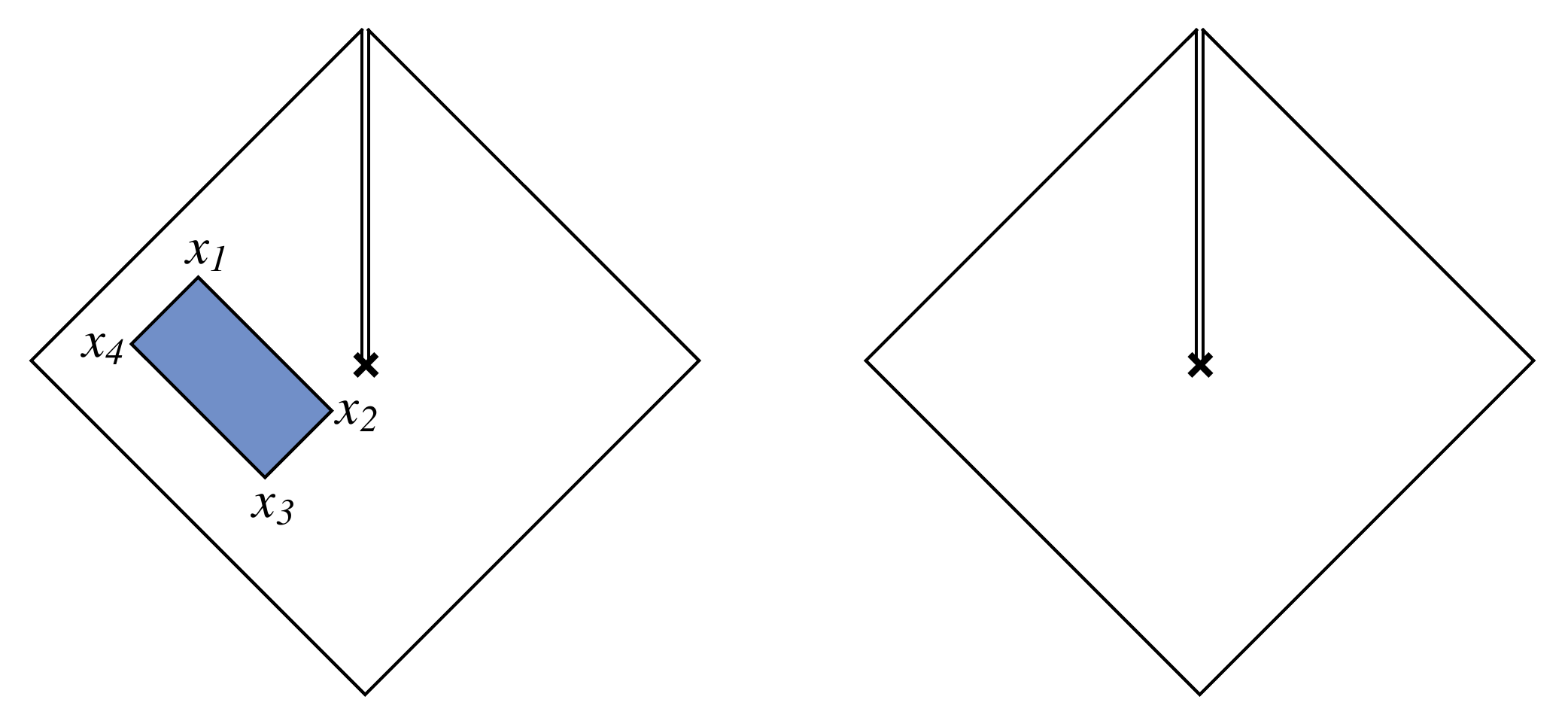}
\caption{A causal interval or causal diamond  not containing the singularity.}
\label{fig:case_i_contour}
\end{figure}

\begin{figure}[t!]
\centering
\includegraphics[
clip=true,
width=\textwidth]
{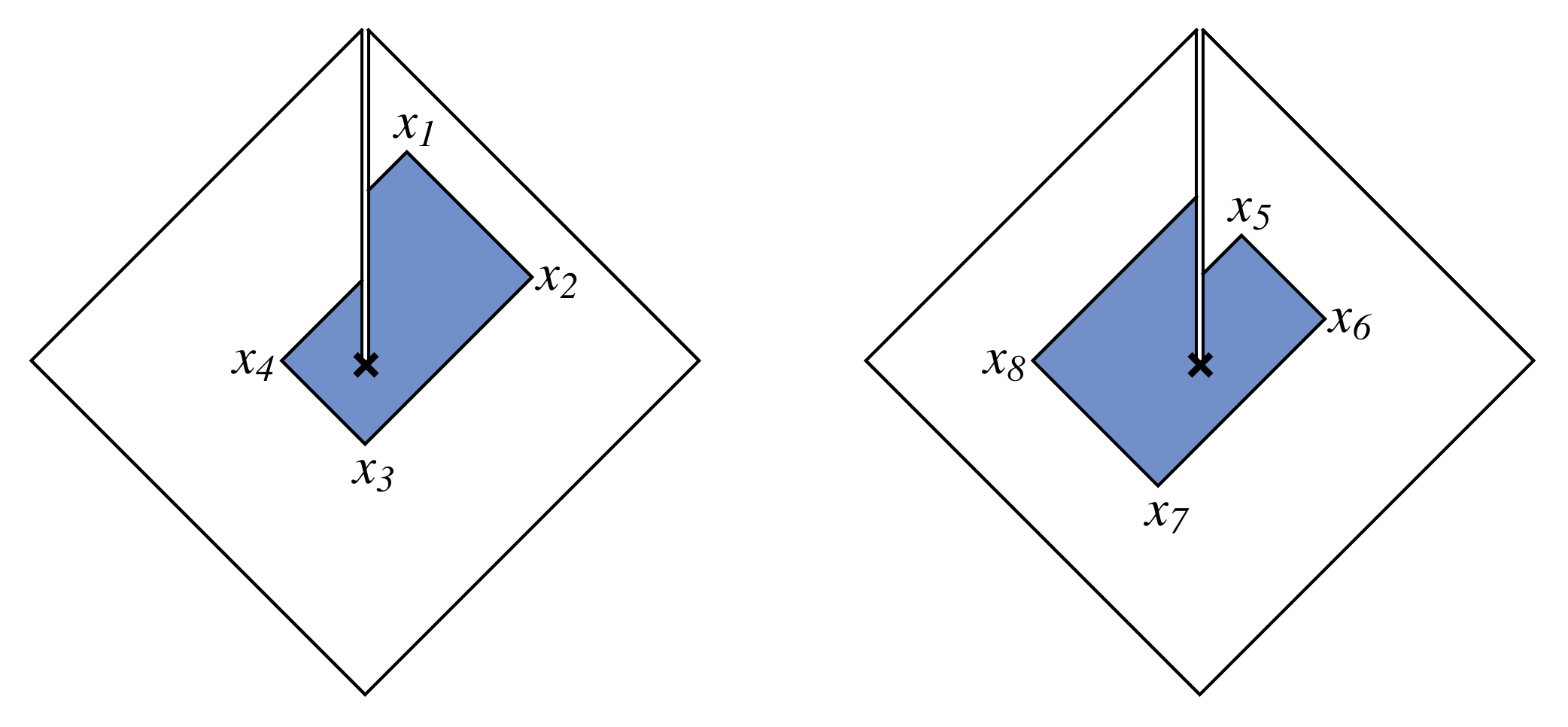}
\caption{Example of a double diamond containing the singularity.}
\label{fig:case_ii_contour}
\end{figure}

We say that function $f$ is a solution of the wave equation if it satisfies 
 \be \label{diamondeom}
 \mathfrak C^{D} f= 0\,,
 \ee
for every causal diamond $D$ that does not contain $x_c$, as illustrated in 
Figure \ref{fig:case_i_contour}, 
and 
 \be \label{ddiamondeom}
 \mathfrak C^{DD} f= 0\,,
 \ee
for each  ``double diamond'', $DD$, whose  interior contains $x_c$ --- like the example shown in Figure 
\ref{fig:case_ii_contour} --- and where the
definition of $\mathfrak C^{DD}$ is the obvious generalisation of (\ref{cee}),
the alternating sum of the values of $f$ at  the vertices of $DD$:
 \be 
 \mathfrak C^{DD} f: =  f(x_1) - f(x_2) + f(x_3) - f(x_4) + f(x_5) - f(x_6) + f(x_7) - f(x_8)\,.
 \ee
The order of the labels of the vertices is clockwise starting from the futuremost
 vertex in region $R_1$ as in Figure \ref{fig:case_ii_contour}.
Note that for each such double diamond, exactly one of its corners lies
in the interior of each of the regions $R_i$ of $\cal{M}$. In the labelling we have chosen,
$x_i \in R_i$. 

It is straightforward to extend this concept of solution
to define a  Green function in $\cal{M}$.
A Green function $G(x,y)$ satisfies
\be \label{ceedee}
 \mathfrak C^{D}_x G(x,y)= -\frac{1}{2} \chi_{D} (y) \,,
 \ee
for every causal diamond, $D$, that does not contain $x_c$, and, in addition,
 \be \label{ceedeedee}
 \mathfrak C^{DD}_x G(x,y)  = -\frac{1}{2} \chi_{DD} (y)\,,
  \ee
for every double diamond, $DD$, surrounding $x_c$. The subscript $x$ on 
the operator $\mathfrak C^{DD}_x$ indicates 
that it
acts on the argument $x$ of $G(x,y)$.


The Hilbert space we are working in is ${L}^2({\cal{M}})$, 
in which members of the same equivalence class differ only
on a set of measure zero.  We say that an element of 
${L}^2({\cal{M}})$ is a solution if it contains a member, $f(x)$, that satisfies the above requirements,~\eqref{diamondeom} and~\eqref{ddiamondeom}. Other members 
of the equivalence class can fail the above conditions but only
on a set of diamonds and double diamonds of measure zero in the space of all diamonds. 

\subsection{A One-Parameter Family of Green Functions}\label{A One-Parameter Family of Green Functions}

In the SJ construction of the quantum theory, the role of the retarded Green function, $G(x,y)$ is its appearance in the Pauli-Jordan function $\Delta(x,y) = G(x,y) - G(y,x)$. The causal structure of the spacetime is imposed on the quantum field theory through the commutation relations $[\phi(x),\phi(y)]=i\Delta(x,y)$, the covariant form of the equal-time canonical commutation relation.  For the field operators to be solutions of the field equations then we also have that $\Delta$ must be a solution to the field equations in both its arguments. We satisfy this condition by requiring that $G(x,y)$ be a Green function in \textit{both} its arguments. 

If a causal interval $[x,y]$ does \textit{not} contain the singularity then 
$[x,y]$ is contained in an open, globally hyperbolic subregion of Minkowski space, and so the retarded Green function $G(x,y)$
 will take its usual Minkowski form, $G(x,y) = G_{\mathsf{Mink}}(x,y).$

\begin{figure}[t]
\centering
\includegraphics[
clip=true,
width=\textwidth]
{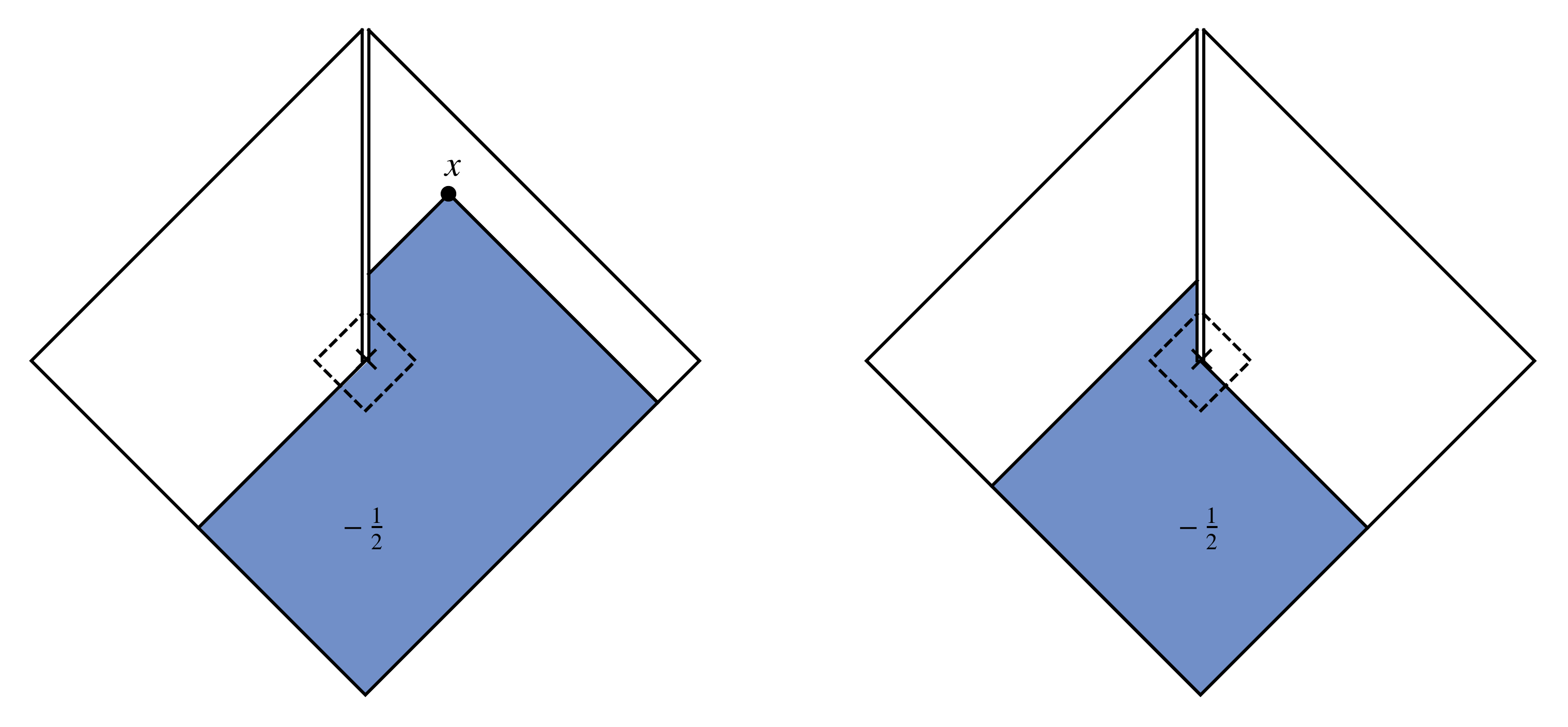}
\caption{The Minkowski retarded Green function $G_{\mathsf{Mink}}(x,y)=-\frac12\chi_{\succ}(x,y)$ in the pair of diamonds, drawn as a function of $y$ for fixed $x$ where 
$x$ is in the causal future of the singularity. The dashed contour corresponds to the boundary of a double diamond, $DD$, centred on the singularity.}
\label{fig:gret-on-pod-false}
\end{figure}

Consider, firstly, $G_{\mathsf{Mink}}(x,y) ) = -\frac{1}{2}\chi_\succ(x,y)$ on the whole of the pair of diamonds
as illustrated in Figure \ref{fig:gret-on-pod-false}.

Choose $x$ to the future of $x_c$ 
and let $DD$ be a double diamond 
around $x_c$ small enough that it does not contain $x$ as shown in Figure  \ref{fig:gret-on-pod-false}. 
In order for $G(x,y)$ to be a Green function in both 
arguments we need it to satisfy, for example,
$\mathfrak C^{DD}_y \,G(x,y)=0$, since $\chi_{DD}(x)=0$. However, 
 $\mathfrak C^{DD}_y\,G_{\mathsf{Mink}}(x,y)=-1/2$. This is reminiscent of the cylinder, in which $G_{\mathsf{Mink}}(x,y)$ does not satisfy Green's equation due to the conjugate points on the cylinder, and this motivates an analogous method of images to find a Green function on the pair of diamonds.  

\begin{figure}[t]
\centering
\includegraphics[
clip=true,
width=1.0\textwidth]
{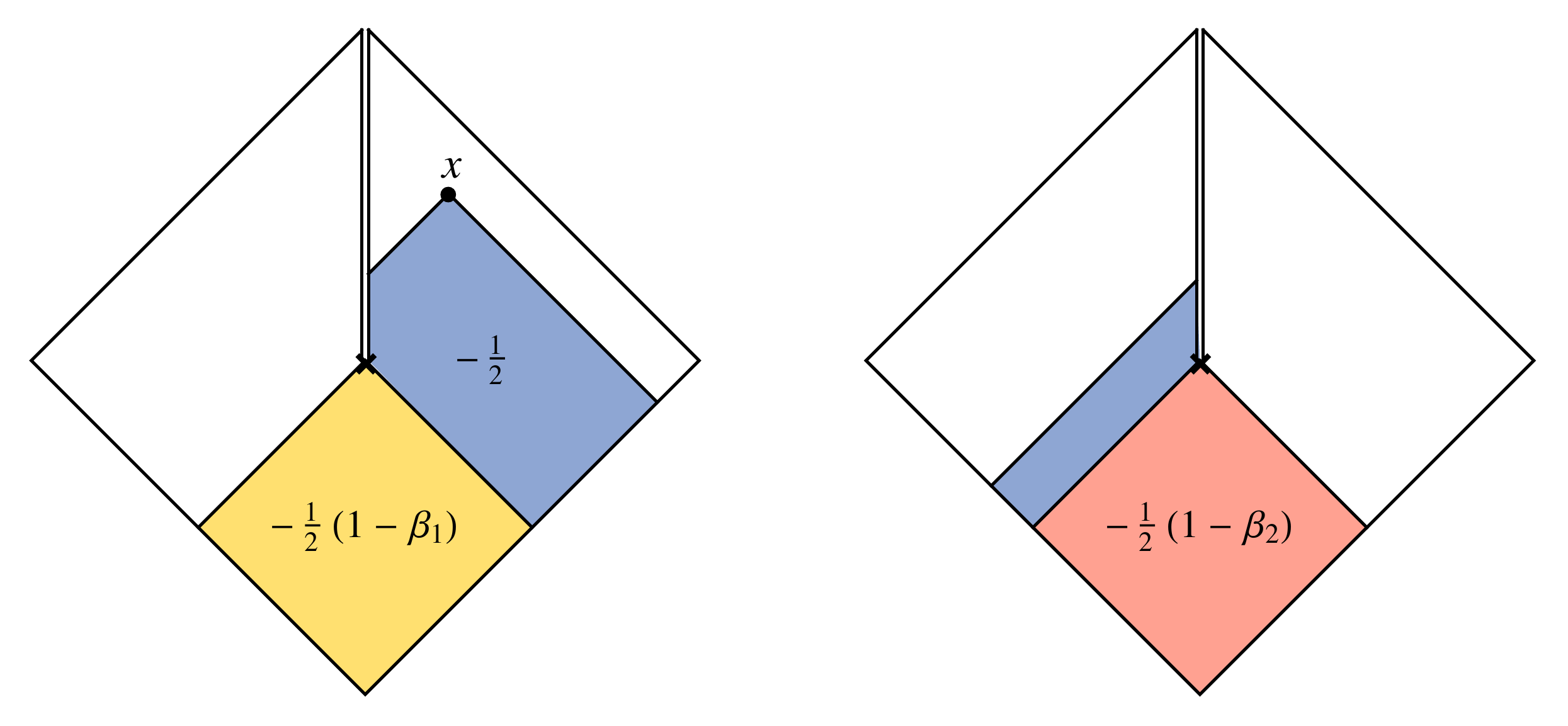}
\caption{An ansatz for the retarded Green function $G(x,y)$ for fixed $x\in R_1$.
If $\beta_1+ \beta_2 = 1$, then ${\mathfrak{C}}_y^{DD} G(x,y) = 0$ for any double diamond,
$DD$, around the singularity.}
\label{fig:gret-on-pod-bs}
\end{figure}

If $x \notin R_1 \cup R_5$  and $y \prec x$ then the interval $[x,y]$ does not 
contain $x_c$ and $G(x,y) = G_{\mathsf{Mink}}(x,y)$. So the only 
cases we need to consider are $x\in R_5$ or $x\in R_1$, and $y \in R_3$ or $y\in R_7$. 

For  $x\in R_1$ let us add to the Minkowski Green function two contributions
from 
an image point at $x_c$, one on diamond $A$ and the other on 
diamond $B$:
\be
\left.G(x,y)\right|_{x\in R_1}=-\frac12\left[\chi_{\succ}(x,y)-\beta_1\chi_3(y)-\beta_2\chi_7(y)\,\right].\label{eq:gret-on-pod-bs}
\ee
 See Figure~\ref{fig:gret-on-pod-bs} for an illustration. Considering a 
double diamond, $DD$, around $x_c$ we find that ${\mathfrak{C}}_y^{DD} G(x,y) = 0$ 
if $\beta_1 + \beta_2 = 1$.  
  
Similarly, for $x\in R_5$, consider the ansatz,
\be
\left.G(x,y)\right|_{x\in R_5}=-\frac12\left[\chi_{\succ}(x,y)-\alpha_1\chi_3(y)-\alpha_2\chi_7(y)\right]\,.\label{eq:gret-on-pod-as}
\ee
Then, $\mathfrak C^{DD}_y G(x,y)=0$ implies $\alpha_1+\alpha_2=1$. 

\begin{figure}[t]
\centering
\includegraphics[
clip=true,
width=1.0\textwidth]
{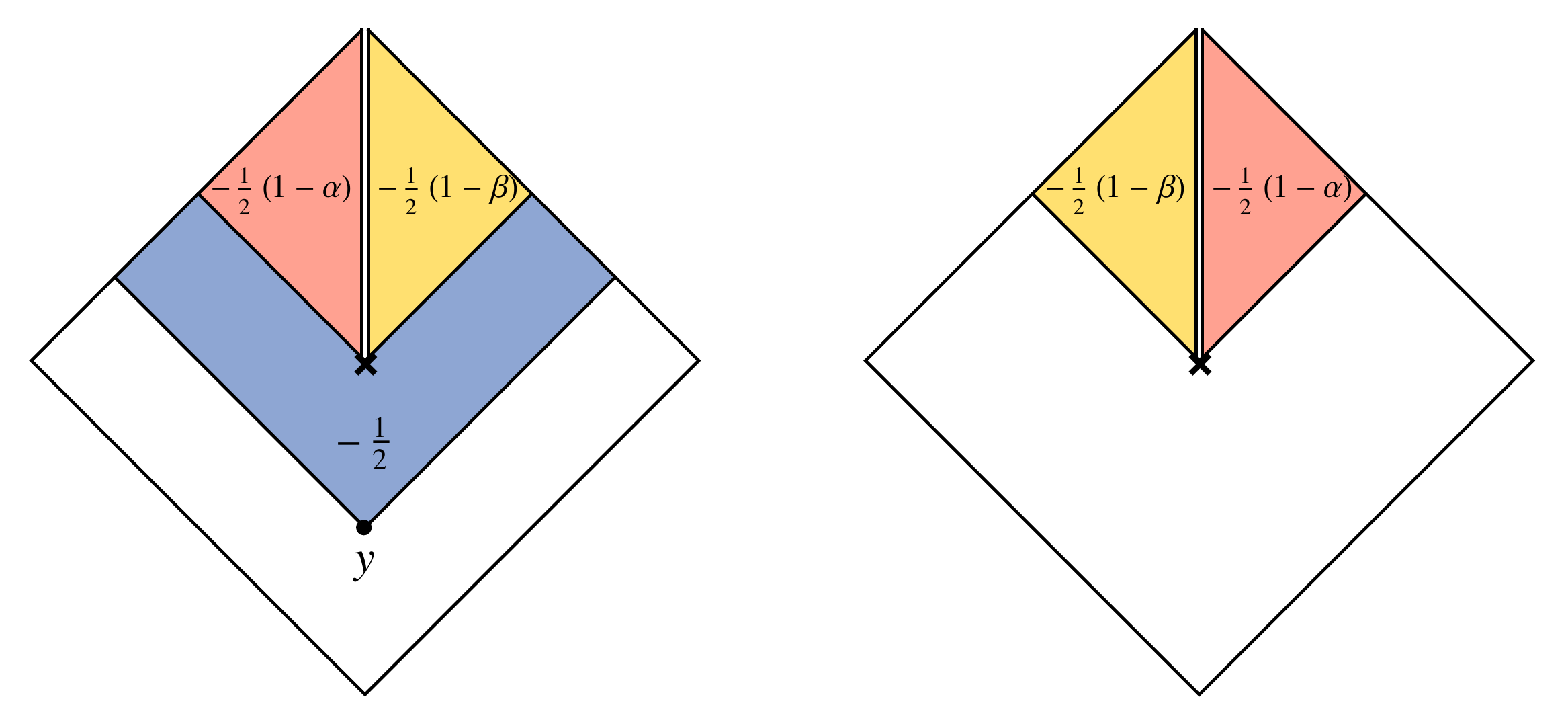}
\caption{The retarded Green function $G_p(x,y)$ for fixed $y$ in $R_3$ as a function of $x$.}
\label{fig:gadv-on-pod-ps}
\end{figure}

This leaves us with a two-parameter family of retarded functions on ${\cal{M}}$, 
with parameters $\alpha:=\alpha_1=1-\alpha_2$ and $\beta:=\beta_1=1-\beta_2$. However, there is a further condition because $G$ is a Green function in its first argument and 
from $\mathfrak C^{DD}_x G(x,y) =0$ for $y \notin DD$ we obtain an additional constraint, $\alpha + \beta = 1$. To see this, fix $y\in R_3$ as in Figure~\ref{fig:gadv-on-pod-ps}, where we have plotted $G(x,y)$ as a function of $x$. If we take a double diamond, $DD$, such that $y \notin DD$, then $\mathfrak C^{DD}_x G(x,y) =-\frac{1}{2}(1-\alpha-\beta)$, and since this must equal $0$ we obtain the constraint $\alpha+\beta=1$.

We are thus left with a one-parameter family of retarded Green functions $G_p(x,y)$ parametrised by $p:= \alpha = 1- \beta$.
The case $p=\frac12$ corresponds to the symmetric case in which the source at $x_c$ is of equal strength in each of the two disconnected pieces of spacetime
that come together or come apart at $x_c$  (see Figures~\ref{fig:gret-on-pod-bs} and~\ref{fig:gadv-on-pod-ps}). 
These additional sources in $G_p(x,y)$ do not by themselves constitute an ``infinite burst in energy''; at this stage they are merely a presage of trouble ahead. 
In order to reach such conclusions, one first has to obtain the quantum state and try to 
compute physical quantities.

\section{Eigenfunctions of the Pauli-Jordan Operator}

\begin{figure}[t!]
\centering
\includegraphics[
clip=true,
width=1.0\textwidth]
{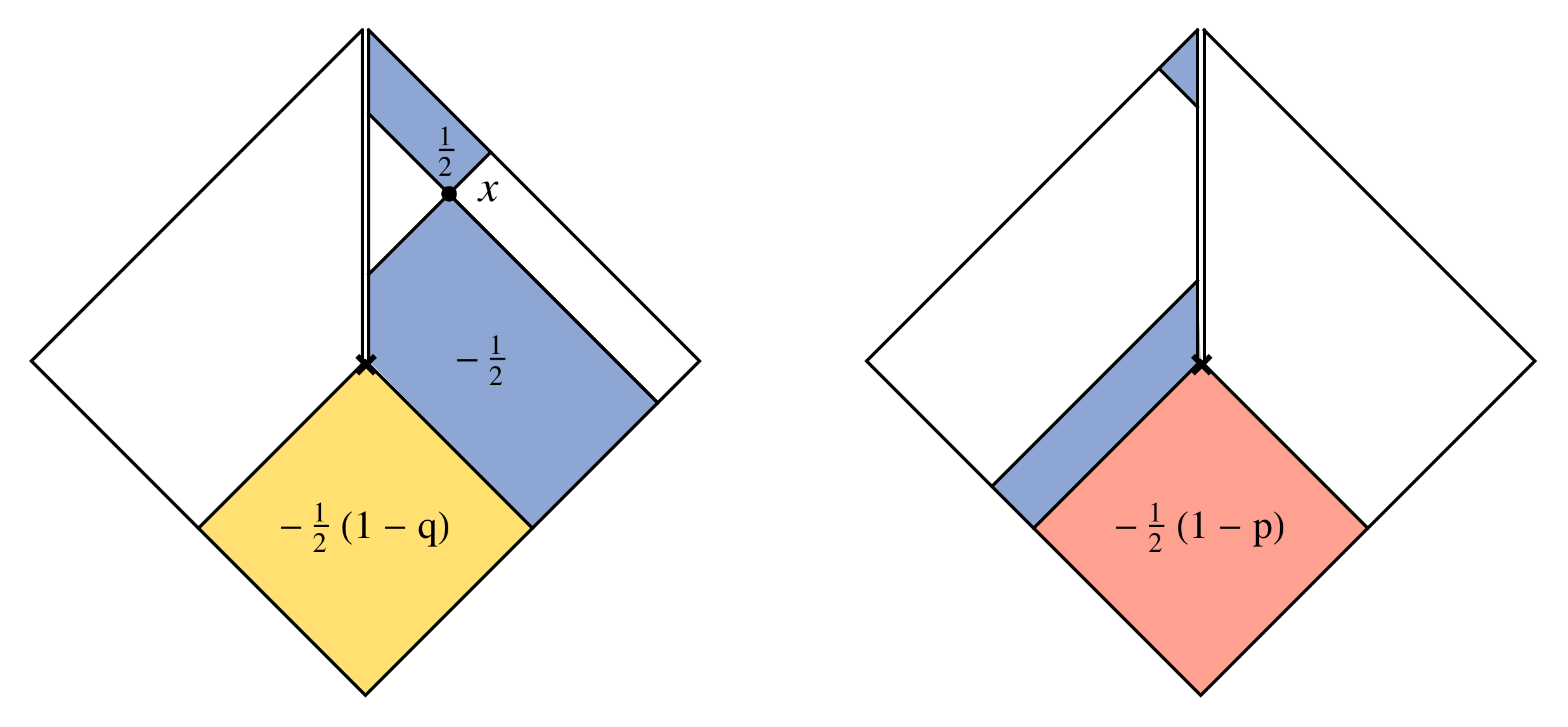}
\caption{The Pauli-Jordan function $\Delta_p(x,y)$ in the pair of diamonds as a function of $y$, with the first argument $x$ fixed  in the causal future of the singularity. Here $q=1-p$.}
\label{fig:pj-on-pod}
\end{figure}

The one-parameter family of retarded Green functions derived in the previous section provides us with a one-parameter family of Pauli-Jordan functions $\Delta_p = G_p - G_p^T$. For an example illustrating its form see Figure~\ref{fig:pj-on-pod}. 
In order to calculate the SJ state our task is now to find the positive part of 
 $i\Delta_p$ and to do that we will solve for the
eigenfunctions of $i\Delta_p$, 
\begin{equation}\label{eigenequation_idelta}
 \int_{\cal{M}} dy\, i\Delta_p(x,y) f(y)=\lambda f(x)\,,
\end{equation}
for $\lambda>0$. 
As mentioned before, the eigenfunctions of $i\Delta_p$ with non-zero eigenvalues come in pairs: the function $f$ with eigenvalue $\lambda>0$, and its complex conjugate, $f^*$, with eigenvalue $-\lambda$. 

Since $i\Delta_p(x,y)$ is a solution in its argument $x$, (\ref{eigenequation_idelta}) 
shows that 
every eigenfunction with non-zero eigenvalue will also be a solution. Indeed, the eigenfunctions with nonzero eigenvalues form a basis
for the space of solutions of the equations of motion on the pair of diamonds. In Appendix~\ref{Zero Eigenvalue Eigenfunctions Are Not Solutions} we show that 
the eigenfunctions with zero eigenvalue --- elements of the kernel of $i \Delta$ ---
are not solutions. 

\subsection{The Norm of the Pauli-Jordan Function}
$i\Delta_p(x,y)$ is a Hilbert-Schmidt integral kernel and its ${L}^2$-norm squared is equal to the sum of the squares of its eigenvalues $\lambda_k$:
\be\label{general_norm_idelta}
\int_{\mathcal{M}} dV_x\int_{\mathcal{M}} dV_y |i\Delta_p(x,y)|^2 = \sum_k\lambda_k^2\,.
\ee
The eigenvalues come in pairs with opposite signs, so this sum is twice the sum of the squares of the positive eigenvalues. The integral on the LHS gives
\begin{gather}
\begin{aligned}
\int_{\mathcal{M}} dV_x\int_{\mathcal{M}} dV_y |i\Delta_p(x,y)|^2 = & \sum_{i,j=1}^8 \int_{R_i} dV_x\int_{R_j} dV_y |i\Delta_p(x,y)|^2
\\
= & 2L^4\left[2-p(1-p)\right]\,.
\label{eq:hsnormpod}
\end{aligned}
\end{gather}
Compare this to the single flat diamond, on which the norm squared of $i\Delta$ equals $2L^4$~\cite{Johnston:2010su}. The relation~\eqref{eq:hsnormpod} is useful because one can check if a given set of eigenfunctions of $i\Delta_p$ is complete: if the eigenvalues sum to less than $2L^4\left[2-p(1-p)\right]$ then there are missing eigenfunctions. Note that the value depends on $p$ so the eigenvalues will be functions of $p$. 
%

\subsection{Isometries and the Pauli-Jordan Function}

The isometries $\mathfrak{P}$ and $\mathfrak{T}$ that generate the isometry group can be represented as operators, 
  $\hat{\mathfrak{P}}$ and $\hat{\mathfrak{T}}$, on the Hilbert space $L^2(\mathcal{M})$.
The action of $\hat{\mathfrak{P}}$ on a function $f(x)$ is given by $\hat{\mathfrak{P}}(f)(x):=f(\mathfrak{P}^{-1}x)$. The action of $\hat{\mathfrak{T}}$ is 
given by  $\hat{\mathfrak{T}}(f)(x):=f^*(\mathfrak{T}^{-1}x)$. 
We can ask if the operators $\hat{\mathfrak{P}}$ and $\hat{\mathfrak{T}}$ commute with $i\Delta_p$. We find that
\begin{gather}
\begin{aligned}
& \hat{\mathfrak{P}}\circ i\Delta_p=i\Delta_{1-p}\circ\hat{\mathfrak{P}}
\\
& \hat{\mathfrak{T}}\circ i\Delta_p=i\Delta_p\circ\hat{\mathfrak{T}}\;,
\end{aligned}
\end{gather}
so that for $p=\frac{1}{2}$ both $\hat{\mathfrak{P}}$ and $\hat{\mathfrak{T}}$ commute with $i\Delta_{\frac{1}{2}}$. This means that $i\Delta_{\frac{1}{2}}$ commutes with the full isometry group.

\subsection{``Copy'' Eigenfunctions}\label{section_single_diamond_modes}
Since we know the SJ modes for the single causal diamond from~\cite{Johnston:2010su}, we can use them as a guide to finding eigenfunctions on the pair of diamonds. In~\cite{Johnston:2010su} it was shown that on the single diamond of area $4{L}^2$, the eigenfunctions of  $i\Delta_{\mathsf{Mink}}(x,y)=-\frac{i}{2}\left[\chi_{\succ}(x,y)-\chi_{\succ}(y,x)\right]$ are linear combinations of positive frequency plane waves and a constant:
\begin{gather}\label{single_diamond_modes}
\begin{aligned}
f_k(u,v) &:= e^{-iku} - e^{-i k v}, & &\quad\textrm{with } k = \frac{n \pi}{L}, \; n = 1, 2, \ldots\\
g_k(u,v) &:= e^{-iku} + e^{-i k v} - 2 \cos(k L), & &\quad\textrm{with } k\in\mathcal{K}
\end{aligned}
\end{gather}
where $\mathcal{K}=\left\{k\in\mathbb{R}\,|\,\tan(kL)=2kL\textrm{ and } k>0\right\}$ and the eigenvalues are $L/k$. The eigenfunctions with eigenvalues $-L/k$ are the complex conjugates of these. Consider now each of these --- $f_k$ and $g_k$ --- modes in turn, 
extended to the pair of diamonds by duplicating the mode onto both diamonds in Figure~\ref{fig:pairofdiamonds}, as if each were a disconnected single diamond. It can be shown that each of these  ``copy modes" on the pair of diamonds is an eigenfunction of $i\Delta_p$, for \textit{any} $p$. The  norm squared of the $f_k$ copy mode on the
 pair of diamonds is
\begin{equation}\label{norm_squared_f_mode}
||f_k||^2:=\int_{-L}^L du_A \int_{-L}^L dv_A {f_k}^* f_k+\int_{-L}^L du_B \int_{-L}^L dv_B {f_k}^* f_k=16L^2\;.
\end{equation}
We define the normalised mode as $\hat{f}_k:=||f_k||^{-1}f_k$. Similarly, we define the normalised mode $\hat{g}_k:=||g_k||^{-1}g_k$, where $||g_k||^2=16L^2\left(1-2\cos(kL) \right)$.

The (positive and negative) eigenvalues of the copy modes sum to $2L^4$, as was shown in~\cite{Johnston:2010su}. Since this is less than the total in~\eqref{eq:hsnormpod}, the copy modes cannot be a complete set.

\subsection{The Other Eigenfunctions}

The form of the remaining eigenfunctions was investigated 
by solving for them in a discrete, finite version of the problem. 
The pair of diamonds was discretised in two different ways, 
with a regular lattice in the coordinates $X$ and $T$, and with causal set sprinklings~\cite{Bombelli:1987aa}. In each case, $i\Delta_p$ is a finite matrix whose indices run over the elements of the lattice or causal set. We solved for the eigenvectors of this matrix numerically and looked for those that did not resemble the $\hat{f}_k$ or $\hat{g}_k$ modes. This led to an ansatz for the extra modes as piecewise continuous functions with the following form: 
\begin{equation}\label{general_form_of_modes}
f(x)=\sum_{i=1}^8\left(a_i e^{-iku}+b_i e^{-ikv}+c_i\right)\chi_i(x)\;,
\end{equation}
where $i$ denotes the region, as shown in Figure \ref{fig:pairofdiamonds},
and the coefficients $a_i$, $b_i$ and $c_i$ are complex. 
When $x$, the argument of $f$,  is in diamond $A$ ($B$) the coordinates $(u,v)$ in (\ref{general_form_of_modes}) are equal to 
$(u_A, v_A)$  ($(u_B, v_B)$).  

The calculations provided evidence that each of the new modes
is odd under interchange of the diamonds, $A\leftrightarrow B$. This implies that $a_i=-a_{i+4}$, $b_i=-b_{i+4}$ and $c_i=-c_{i+4}$ for $i=1,...,4$. The calculations also showed that the
modes are discontinuous across the past and future directed null lines from the origin on both diamonds. 

All the non-zero eigenvalue eigenfunctions of $i\Delta_p$ are solutions of the
wave equation. Using~\eqref{diamondeom} for a diamond straddling the boundary between two regions, gives conditions on the constants:
\begin{equation}\label{conditions_bewteen_constants}
a_1=-a_4\, , \; a_2=a_3\, , \; b_1=b_2\, , \; b_3=b_4\;.
\end{equation}
The above conditions leave us with $8$ complex parameters $\lbrace a_1,a_2,b_1,b_3,c_1,c_2,c_3,c_4\rbrace$. These, and the allowed values of $k$, 
are fixed by the eigenvalue equation for $i\Delta_p$. In the following sections we will only discuss the eigenfunctions with positive eigenvalues unless otherwise stated. The eigenvalues are given in terms of $k$ by $\lambda_k = L/k$.

\subsection{ $p = \frac{1}{2}$} \label{disconthalf}
 In this case $k>0$ satisfies
\be\label{phalfevaleqn}
\left(2 + \left(kL\right)^2\right) \cos \left(kL\right) + 2 kL \sin \left(kL\right) - 2 = 0\,.
\ee
  The eigenvalue corresponding to 
each solution of this equation is degenerate and  there are two modes 
with that eigenvalue, one for which $a_1 = b_1$ and one for which 
$a_1 = -b_1$. 

\subsubsection{ $a_1=b_1$}

\begin{figure}[h!]
\centering
\includegraphics[
clip=true,
width=0.7\textwidth]
{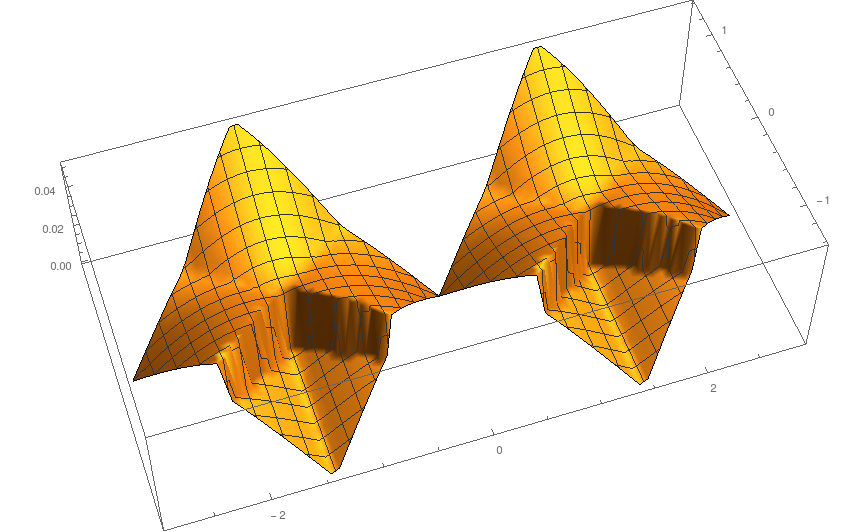}
\hspace{0.1\textwidth}
\includegraphics[
clip=true,
width=0.7\textwidth]
{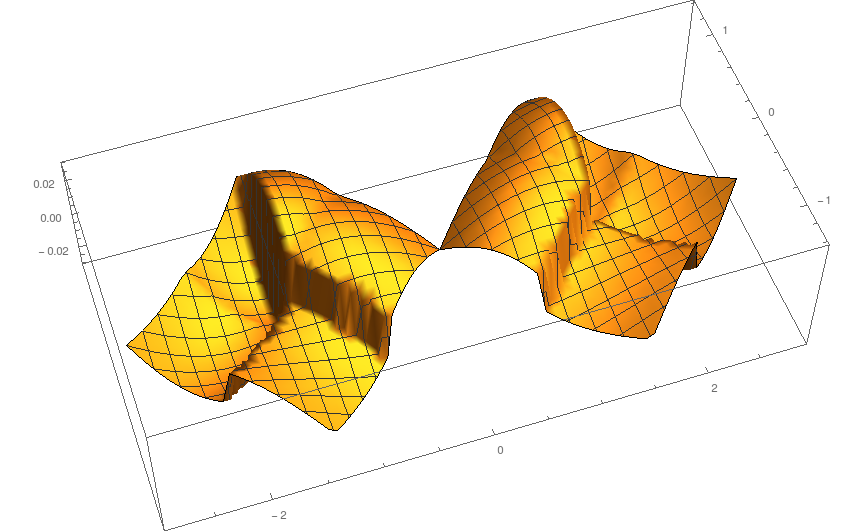}
\hspace{0.01\textwidth}
\includegraphics[
clip=true,
width=0.7\textwidth]
{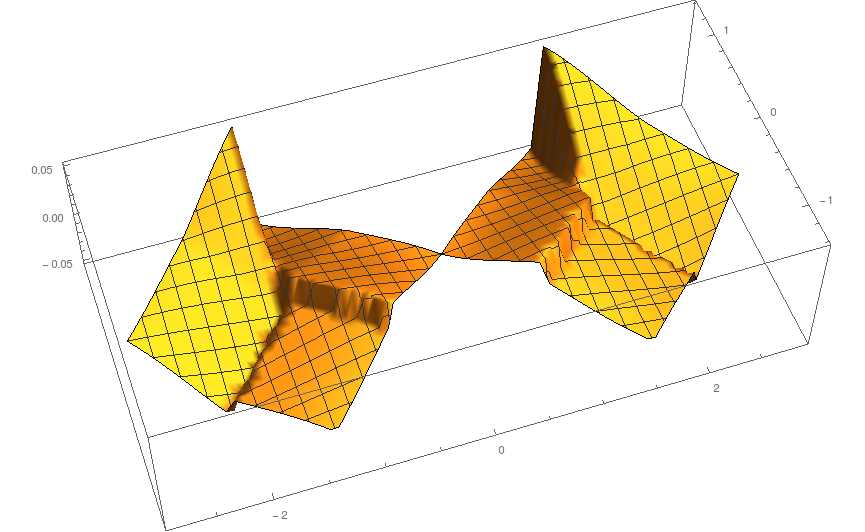}
\caption{The $\hat{f}_k^{(\frac{1}{2})}$ mode for the lowest $k$ satisfying~\eqref{phalfevaleqn}. On the top we have plotted the absolute value of the mode across the pair of diamonds. In the middle we have plotted its real part, and at the bottom its imaginary part. The discontinuity across the line of $X=0$ for $T>0$ is not a discontinuity in the mode. It is simply a consequence of how we have set up the identifications on the pair of diamonds.}
\label{fig:p_half_mode}
\end{figure}

 The coefficients are
\begin{equation}\label{coeffs_p_half_a_b}
\begin{aligned}
a_1=b_1&=kL+2i\\
a_2=-b_3 &= i kL \cot \left(\frac{kL}{2}\right) e^{-i kL}\\
c_1 &=-2 i \left(1+e^{-i kL}\right)\\
c_2 = -c_4 &=-\frac{2}{kL}\left(1-ikL-e^{-i kL}\right)\\
c_3&=0\;.
\end{aligned}
\end{equation}
We denote the mode with these coefficients as $f_k^{(\frac{1}{2})}$. The norm-squared of this mode is
\be
||f_k^{(\frac{1}{2})}||^2=8\frac{L}{k}\left(8 kL + 4 kL \cos \left(kL\right)+ \left(kL\right)^3 \csc^2\left(\frac{kL}2\right) - 8 \sin \left(kL\right)\right)\;.
\ee
The mode that is normalised under the ${L}^2$ inner product is then $\hat{f}_k^{(\frac{1}{2})}:=||f_k^{(\frac{1}{2})}||^{-1}f_k^{(\frac{1}{2})}$. The lowest $k$ mode is plotted in Figure \ref{fig:p_half_mode}. 

\subsubsection{ $a_1=-b_1$}

The coefficients are 
\begin{equation}\label{coeffs_p_half_a_minus_b}
\begin{aligned}
a_1=-b_1&=ikL\cot\left(\frac{kL}{2}\right)e^{ikL}\\
a_2=b_3 &= k-2i \\
c_1 &=0\\
c_2 = c_4 &=-\frac{2}{kL}\left(1+ikL-e^{ikL}\right)\\
c_3&=2i(1+e^{ikL})\;.
\end{aligned}
\end{equation}
A mode with these coefficients will be denoted as $g_k^{(\frac{1}{2})}$. The norm-squared is $||g_k^{(\frac{1}{2})}||=||f_k^{(\frac{1}{2})}||$
and the normalised mode is $\hat{g}_k^{(\frac{1}{2})}:=||g_k^{(\frac{1}{2})}||^{-1}g_k^{(\frac{1}{2})}$. The $\hat{f}^{(\frac{1}{2})}$ modes and the $\hat{g}^{(\frac{1}{2})}$
are orthogonal. The phase of $\hat{g}_k^{(\frac{1}{2})}$ was chosen such that $\hat{\mathfrak{T}}(\hat{f}_k^{(\frac{1}{2})})=\hat{g}_k^{(\frac{1}{2})}$.

$i\Delta_\frac12$ commutes with the isometry group $D_4$  and for each $k$ the $2$-dimensional eigensubspace of $i\Delta_\frac12$, spanned by $\{g_k^{(\frac{1}{2})}, 
f_k^{(\frac{1}{2})}\}$, carries the $2$-dimensional irreducible representation of $D_4$.

In Appendix \ref{appendix_sum_of_square_evals} we verify that these, and the $\hat{f}_k^{(\frac{1}{2})}$ modes, are indeed all the extra modes. That is, we show that the sum of the squares of the eigenvalues (both the positive and negative values) for the modes $\hat{f}_k$, $\hat{g}_k$, $\hat{f}_k^{(\frac{1}{2})}$ and $\hat{g}_k^{(\frac{1}{2})}$ is
\begin{equation}\label{sum_evals_p_half}
\sum_{\text{all modes}} {\lambda_k}^2 = \frac{7L^4}{2}\,.
\end{equation}
The right side of~\eqref{sum_evals_p_half} agrees with~\eqref{eq:hsnormpod} when $p=\frac{1}{2}$.

\subsection{$p\neq \frac{1}{2}$} \label{discontmodes}

We start with the ansatz for a mode 
(\ref{general_form_of_modes})
with $a_{i+4} = -a_i$, $b_{i+4} = - b_i$, $c_{i+4} = - c_i$ for $i = 1,\dots 4$, 
$a_1 = -a_4$, $a_2 = a_3$, $b_1 = b_2$ and $b_3 = b_4$, as before. 
For $p\ne \frac{1}{2}$ we expect to see a dependence on $p$ in the coefficients. With this ansatz one can show that each eigenvalue, $\lambda_k$,
satisfies one of two  possible equations:
\begin{equation}\label{general_p_eigenval_eqns}
\left(\left(kL\right)^2+2\right) \cos\left(kL\right)+kL (2\pm kL (1-2 p)) \sin\left(kL\right)-2=0\,,
\end{equation}
where $k = \frac{L}{\lambda_k}$,
This is consistent with the $p=\frac{1}{2}$ case as the above two equations become \eqref{phalfevaleqn} when $p=\frac{1}{2}$. By using the ansatz~\eqref{general_form_of_modes}, and by using \eqref{general_p_eigenval_eqns} to simplify the resulting equations we find that the coefficients are
\begin{gather}\label{general_p_coeffs}
\begin{aligned}
a_1&= e^{i kL} kL \left\{ i (1+kL (kLp+i)) +e^{i kL} \left[kL (2-kL (kL+i) (p-1))-3 i\right]\right.\\ 
&\qquad\quad\left.+i e^{2 i kL} \left[3-kL (kL (p-1)-i)\right]+e^{3 i kL} \left[(kL)^2(kL p+i (p-2))-i\right]\right\}\\
a_2&=kL \left\{i-kL (2+kL (kL+i) p) + e^{i kL} \left[kL (3+i kL (p-1))-3 i\right] \right.\\
&\left.\qquad\quad+ e^{2 i kL} \left[(kL)^2 (i(p+1)+kL (p-1))+3 i\right] -i e^{3 i kL} \left[1+kL (kL p-i)\right]\right\}\\
b_1&=e^{i kL} kL \left\{i-(kL+i kL (p-1))-e^{i kL} \left[3 i-kL (2+kL (kL+i) p)\right]\right.\\
&\qquad\quad\left.+e^{2 i kL} \left[(kL)(ikLp-1)+3 i\right]-e^{3 i kL} \left[(kL)^2(p+1+kL(p-1))+i\right]\right\}\\
b_3&=kL \left\{kL (2-kL (kL+i) (p-1))-i+i e^{i kL} \left[3+kL (kL p+3 i)\right]\right.\\
&\qquad\quad\left.+e^{2 i kL} \left[i (kL)^2 ((p-2)+kL p)-3 i\right]+e^{3 i kL} \left[kL(1-i kL(p-1))+i\right]\right\}\\
c_1&=2 e^{i kL} \left(e^{i kL}+1-ikL\right) (e^{ikL}-1) \left(\left((kL)^2+2\right) \cos (kL)+2 kL \sin (kL)-2\right)\\
c_2&=2 e^{2 i kL} kL \sin(kL) \left[kL (i kL (1-2 p) + 2) \sin(kL)-\left((kL)^2+2\right) \cos (kL)+2\right]\\
c_3&=(kL)^2 (2 p-1)\left(1-e^{i kL}\right)^2 \left(1+e^{i kL}\right) \left(e^{i kL} (1-i kL)-1\right)\\
c_4&=2 e^{2 i kL} kL \sin(kL) \left[kL (i kL (1-2 p) - 2) \sin(kL) + \left((kL)^2+2\right) \cos (kL)-2\right].
\end{aligned}
\end{gather}

A mode with these coefficients and $k$ satisfying~\eqref{general_p_eigenval_eqns} with the ``$+$" sign will be denoted as $f_k^{(p)}$. Likewise, for the ``$-$" sign we call the mode $g_k^{(p)}$. The $p\neq \frac{1}{2}$ case differs from the $p=\frac{1}{2}$ case in that the coefficients have the same form in terms of $k$ for both the $f_k^{(p)}$ and $g_k^{(p)}$ modes. The $f_k^{(p)}$ and $g_k^{(p)}$ modes still have different coefficients, though, because the allowed
values of $k$ are different
 as they come from~\eqref{general_p_eigenval_eqns} with either the ``$+$" or ``$-$" sign.

In Appendix \ref{appendix_sum_of_square_evals} we verify that these two sets of modes, together with the  $f_k$ and $g_k$ copy modes, are all the eigenfunctions of $i\Delta_p$ with positive eigenvalues. There we show that the sum of the squares of the eigenvalues for all the modes agrees with the right hand side of~\eqref{eq:hsnormpod}. That is, 
\begin{equation}\label{sum_evals_p}
\sum_{\text{all modes}} {\lambda_k}^2 = 2L^4\left(2-p(1-p) \right)\;.
\end{equation}

The norm-squared for either mode has the same form in terms of $k$, and is
\begin{equation}
\begin{aligned}
\label{general_p_norm}
&||f_k^{(p)}||^2=||g_k^{(p)}||^2=32 k^5L^7 (1 - 2 p)^2 \sin^2(kL) \\
&\qquad\times\left[kL (3 + (kL)^2 - 2 \cos(kL) - \cos(2kL)) + 4 (\cos(kL) -1) \sin(kL)\right]\;.
\end{aligned}
\end{equation}

We define the normalised modes $\hat{f}_k^{(p)}:=||f_k^{(p)}||^{-1}f_k^{(p)}$ and   $\hat{g}_k^{(p)}:=||g_k^{(p)}||^{-1}g_k^{(p)}$. Both these modes tend to the $\hat{f}_k^{(\frac{1}{2})}$ mode in the $p\rightarrow\frac{1}{2}$ limit. That is, 
\be
\lim_{p\rightarrow\frac{1}{2}}\hat{f}_k^{(p)}=\lim_{p\rightarrow\frac{1}{2}}\hat{g}_k^{(p)}=\hat{f}_k^{(\frac{1}{2})}. 
\ee
The $\hat{g}_k^{(\frac{1}{2})}$ mode appears as an entirely new eigenfunction (in the sense that the coefficients for this mode have a different form in terms of $k$) only when $p=\frac{1}{2}$.

\section{Energy momentum in the SJ State}

Knowing the complete set of positive eigenvalue eigenfunctions of  $i\Delta$ means that one knows the SJ state since its Wightman function can be expressed as the sum (\ref{eq:W}) over these eigenfunctions. For each $p$, we have found this complete set and so we have the SJ state. We can now turn to studying what physical properties this SJ state has. Sorkin argues that, ultimately, quantum field theory should be based on the path integral and will not be able to be fully self-consistent except within a theory of quantum gravity in which the effect of quantum matter on spacetime itself is taken into account  \cite{Sorkin:2011pn}. Quantum gravity and the interpretation of path integral quantum theory are works in progress, so
we will proceed here by seeing what can be gleaned by investigating the expectation value of the energy momentum tensor, $T_{\mu\nu}$. In order to calculate this expectation value one can regulate the divergence of the
 Wightman function and its derivatives in the coincidence limit 
 using point splitting and subtraction of the corresponding quantity in the ``same'' theory in Minkowski spacetime, if the state has the Hadamard property. 
 Fewster and Verch \cite{Fewster}
 showed that the SJ state in a finite slab of a cosmological spacetime with closed 
 spatial sections generically is not Hadamard. It seems likely that the SJ state in the pair of diamonds is also not
 Hadamard since the SJ state for the single diamond is not
  ~\cite{Yasaman:privatecom}. It is possible that the SJ states in the single diamond and pair of diamonds can be rendered Hadamard by a smoothing of the boundary of the diamond ~\cite{Brum:2013bia} and it is an open question whether the Hadamard property should be considered to be physically significant when quantum gravity
suggests that the differentiable manifold structure of spacetime breaks down at the Planck scale.  Here we will simply ignore this question and provide heuristic evidence that an infinite burst of energy along the
lightcones from the singularity will be present in the SJ state. 

A creation and annihilation operator can be assigned to each mode and
the field operator can be written as a sum over modes ~\cite{Johnston:2009fr,Sorkin:2011pn,Afshordi:2012jf}
\begin{equation}\label{field_operator}
\phi(x)=\sum_\veca \sqrt{\lambda_\veca}\left(\mathfrak u_\veca(x) a_\veca + \mathfrak{u}_\veca^*(x) a_\veca^{\dagger} \right)\,,
\end{equation}
where  $\{\mathfrak u_\veca\}$ are the orthonormal eigenfunctions of $i\Delta_p$ with  positive eigenvalues $\lambda_\veca$ and  $[a_\veca, a^\dagger_{\vecb}] = \delta_{\veca\vecb}$ and  $[a_\veca, a_{\vecb}] = 
[a^\dagger_\veca, a^\dagger_{\vecb}] = 0$.
The SJ state, $\big| 0_{(p)} \big>$, is then the state that is annihilated by $a_\veca$  for all $\veca$.  For each $p$ there is an inequivalent quantum theory.

The operator for the stress energy of the massless field is
\begin{equation}\label{stress_energy}
T_{\alpha\beta}=\phi_{,\alpha}\phi_{,\beta}-\frac{1}{2}\eta_{\alpha\beta}\eta^{\lambda\sigma}\phi_{,\lambda}\phi_{,\sigma}\;,
\end{equation}
in Cartesian $(T,X)$ coordinates in which the metric locally is the Minkowski 
metric, $\eta_{\alpha \beta}$. We can construct the operator for the energy on the future (or past) null boundary of the pair of diamonds by integrating $T_{\alpha\beta}\xi^{\beta}$ across the surface, where $\xi^{\alpha}$ is the Killing vector $\partial/\partial T$. Let $N_+$ be the future null boundary of $\mathcal{M}$. The energy operator for this boundary is
\begin{equation}
E_+:=\int_{N_+}d\Sigma^{\alpha}T_{\alpha\beta}\xi^{\beta}\;.
\end{equation}
Using~\eqref{stress_energy} and converting to light-cone coordinates, this becomes
\begin{gather}\label{energy_bottom_surface}
\begin{aligned}
E_+=\frac{1}{\sqrt{2}}\Bigg( & \int_{-L}^{L}du_A (\phi_{,u_A})^2\Big|_{v_A=L}+\int_{-L}^{L}dv_A (\phi_{,v_A})^2\Big|_{u_A=L}
\\
&\int_{-L}^{L}du_B (\phi_{,u_B})^2\Big|_{v_B=L}+\int_{-L}^{L}dv_B (\phi_{,v_B})^2\Big|_{u_B=L}\Bigg)\;,
\end{aligned}
\end{gather}
where the first (second) line comes from integrating over the part of the surface on diamond $A$ ($B$). We can similarly define the energy operator $E_-$ for the past null boundary $N_-$.

\subsection{$p=\frac{1}{2}$}

Using the expansion for the field operator in the SJ modes gives the formal expression
\begin{equation}\label{energy_density_p_half}
\begin{aligned}
\big< 0_{(\frac{1}{2})}\big| E_+ \big| 0_{(\frac{1}{2})}  \big>  =  & \sqrt{2}L\int_{-L}^Ldu \bigg(\sum_kk^{-1}\partial_u \hat{f}_k \partial_u \hat{f}_k^*+\sum_{k\in\mathcal{K}}k^{-1}\partial_u \hat{g}_k \partial_u \hat{g}_k^*
\\
+ & \sum_kk^{-1}\Big(\partial_u \hat{f}_k^{(\frac{1}{2})} \partial_u \hat{f}_k^{(\frac{1}{2})*}+\partial_u \hat{g}_k^{(\frac{1}{2})} \partial_u \hat{g}_k^{(\frac{1}{2})*}\Big)\bigg)\bigg|_{v=L}
\\
+ &\, ( u  \leftrightarrow v)\;,
\end{aligned}
\end{equation}
where the $(u,v)$ coordinates refer to the light-cone coordinates on either diamond, as both diamonds give the same result.
In the first sum in (\ref{energy_density_p_half}) $k=\frac{n\pi}{L}$, where $n\in\mathbb{N}$, and the third sum runs over the positive roots of~\eqref{phalfevaleqn}.

This expression~\eqref{energy_density_p_half} involves {products} of derivatives of the discontinuous SJ modes so it is not rigorously defined. However, we see that
as the discontinuities are along the past and future directed light rays from $x_c$,
the integrals along the $v=L$ and $u=L$ lines in~\eqref{energy_density_p_half} have integrands that contain squared Dirac-delta functions located at $u=0$ and $v=0$ respectively. The same situation also arises in the expectation value of $E_-$. This squared Dirac-delta 
divergence was found in previous works on the trousers, although here the divergence 
is along both the past and the future lightcones of the singularity, while in previous work the 
divergence only appears in the future. We now check that the delta-function squared terms have positive coefficients.

Restricting attention to the integral over the $v=L$ line, a mode has the following form:
\begin{equation}\label{p_half_general_mode_stress_energy}
\left(\Theta(u)a_1+ \Theta(-u)a_2\right)e^{-iku}+b_1e^{-ikL}+\left(\Theta(u)c_1+ \Theta(-u)c_2\right)\;,
\end{equation}
up to some normalisation constant, and the coefficients are given by \eqref{coeffs_p_half_a_b} or~\eqref{coeffs_p_half_a_minus_b}. 

Taking a $u$ derivative of the mode in~\eqref{p_half_general_mode_stress_energy} and ignoring the parts with no $\delta$-function dependence we get
\begin{equation}\label{p_half_delta_term}
\delta(u)\left( (a_1-a_2)e^{-iku}+c_1-c_2 \right)\;.
\end{equation}
Each of the terms  $\partial_u \hat{f}_k^{(\frac{1}{2})} \partial_u \hat{f}_k^{(\frac{1}{2})*}$ and $\partial_u \hat{g}_k^{(\frac{1}{2})} \partial_u \hat{g}_k^{(\frac{1}{2})*}$ in the sum in~\eqref{energy_density_p_half}  gives a contribution to the energy equal to $\delta(0)$ times a 
positive coefficient if the complex number $(a_1$ $-\;a_2$ $+\;c_1$ $-\;c_2)$ is non-zero. Using~\eqref{coeffs_p_half_a_b}, and the eigenvalue equation~\eqref{phalfevaleqn}, we find that this complex number is zero for the $\hat{f}_k^{(\frac{1}{2})}$ mode and is non-zero for $\hat{g}_k^{(\frac{1}{2})}$.

A similar conclusion can be drawn for the integral over the $u=L$ line. There, the $\hat{f}_k^{(\frac{1}{2})}$ mode doesn't contribute whilst the $\hat{g}_k^{(\frac{1}{2})}$ mode does. For the expectation value of $E_-$ the situation is reversed --- the $\hat{g}_k^{(\frac{1}{2})}$ mode doesn't contribute while the $\hat{f}_k^{(\frac{1}{2})}$ mode does. Therefore, on both the past and future null boundaries of $\mathcal{M}$ there appears to be a divergence in the energy. This divergence implies that the QFT in curved spacetime approximation — in which back reaction on the spacetime is ignored — must break down. It could be a signal that the trousers topology change cannot occur at all but at the very least it means that the spacetime cannot be approximated by the flat geometry we have been working with. 

\subsection{$p\neq \frac{1}{2}$}

The expectation value of $E_+$ in the SJ state is 
\begin{equation}\label{energy_density_p}
\begin{aligned}
\big< 0_{(p)}\big| E_+ \big| 0_{(p)}  \big>  =  & \sqrt{2}L\int_{-L}^Ldu \bigg(\sum_kk^{-1}\partial_u \hat{f}_k \partial_u \hat{f}_k^*+\sum_{k\in\mathcal{K}}k^{-1}\partial_u \hat{g}_k \partial_u \hat{g}_k^*
\\
+ & \sum_kk^{-1}\partial_u \hat{f}_k^{(p)} \partial_u \hat{f}_k^{(p)*}+\sum_kk^{-1}\partial_u \hat{g}_k^{(p)} \partial_u \hat{g}_k^{(p)*}\bigg)\bigg|_{v=L}
\\
+ &\, (u  \leftrightarrow v)\;,
\end{aligned}
\end{equation}
where the first two sums are over the same values of $k$ as those in~\eqref{energy_density_p_half}, and the last two sums are over the solutions of~\eqref{general_p_eigenval_eqns} with the ``$+$" and ``$-$" signs respectively. 

For $p\neq 1$ and $\neq 0$ one finds that, on all parts of the null boundaries, both $\hat{f}_k^{(p)}$ and $\hat{g}_k^{(p)}$ modes contribute $\delta(0)$ terms to the expectation value of $E_+$ and $E_-$. However, when $p=0$ there is no divergence on the lefthand segments of $N_+$ and $N_-$ \textit{i.e.} $u=L$ and $v=-L$, respectively. For $p=1$ there is no divergence from the righthand segments of $N_+$ and $N_-$, \textit{i.e.} the lines  $v=L$ and $u=-L$, respectively.

\section{From the Pair of Diamonds to the Infinite Trousers}\label{From the Pair of Diamonds to the Infinite Trousers}

In this section we provide further evidence that the divergence in energy is located
along the past and future lightcones of the singularity by
examining the infinite limit of the pair of diamonds. This allows 
us better to compare the SJ state to scalar QFT in 1+1 Minkowski spacetime. Specifically, we take $L\rightarrow\infty$ in the pair of diamonds to get two copies of Minkowski spacetime with trousers-type identifications along the positive time axes. We call this double sheeted Lorentzian spacetime the \textit{infinite trousers}. The two  planes are labelled $A$ and $B$ in the same way as the pair of diamonds. The conformal compactification of the infinite trousers is the pair of diamonds. The retarded Green function is the same function, $i\Delta_p$ as in the pair of diamonds. 

We take an appropriate limit of the eigenfunctions of $i\Delta_p$  and compare them with the usual modes of Minkowski spacetime. Strictly, we are leaving the finite spacetime volume regime in which the SJ formalism is defined. Nevertheless,
we can renormalise the modes in order that they have a sensible limiting form and display the 
usual feature of the passage from a finite box to an infinite spacetime, namely the 
transition from a countable set of modes to an uncountable, delta-function normalised set. 


Consider first the $\hat{f}_k$ copy modes. We
define  $f_{n}^L:= \frac{L}{\sqrt{\pi k}}\hat{f}_k$ where natural number $n$ labels the 
eigenvalues in increasing order, in this case via the simple relationship
 $k = \frac{n \pi}{L}$. 
For  each \textit{real}  number $k>0$ and each value of $L$, we can find an integer $n_{k,L}$
such that $\lim_{L\to \infty} \frac{\pi }{L}n_{k,L}= k$. Indeed $n_{k,L} = 
 \lfloor \frac{L k}{\pi} \rfloor$ will do the job. 

Then,  in the limit $L\rightarrow\infty$, for each real $k>0$ we define the infinite trousers copy mode $\tilde{f}_k:=\lim_{L\rightarrow \infty}f_{n_{k,L}}^L=\frac{1}{\sqrt{16\pi k}}\left(e^{-iku}-e^{-ikv} \right)$, where coordinates $u$ and $v$ here
are light-cone coordinates on the infinite trousers. 

Considering the $\hat{g}_k$  modes, we
define $g_{n}^L:=\frac{L}{\sqrt{\pi k}}\hat{g}_k$ where 
$n$ labels the discrete eigenvalues $k_n$  satisfying $\tan(kL)=2kL$ in increasing order. Now there is no simple relationship between $n$ and eigenvalues $k_n$ but 
 $k_n\to (n+\frac{1}{2})\frac{\pi}{L}$ as $n \to \infty$. So, again, for each real $k>0$ and all
 values of $L$ there exist integers
  $n_{k, L}$ such that $\lim_{L\to \infty} (n_{k,L}+ \frac{1}{2})\frac{\pi }{L}= k$.
Then,  in the limit $L\rightarrow\infty$, for each real $k>0$ we define the infinite trousers copy mode  $\tilde{g}_k:=\lim_{L\to \infty}g_{n_{k,L}}^L=\frac{1}{\sqrt{16\pi k}}\left(e^{-iku}+e^{-ikv} \right)$. 

\subsection{The Discontinuous Modes in the Infinite Trousers}\label{The Extra Modes in the Infinite Trousers}

The discontinuous modes in the infinite trousers are odd under interchange of 
the two sheets and, using the same limiting procedure as above applied to the modes
${\hat{f}}^{(p)}$ and ${\hat{g}}^{(p)}$ from section \ref{discontmodes}, we obtain 
\begin{equation}\label{infinite_p_modes_general_form}
\begin{aligned}
& \tilde{f}_k^{(p)}(x)=\frac{1}{\sqrt{16\pi k}}\sum_{i=1}^8\left(a^f_i e^{-iku}+b^f_i e^{-ikv}\right)\chi_i(x)
\\
& \tilde{g}_k^{(p)}(x)=\frac{1}{\sqrt{16\pi k}}\sum_{i=1}^8\left(a^g_i e^{-iku}+b^g_i e^{-ikv}\right)\chi_i(x)\;,
\end{aligned}
\end{equation}
respectively, where the coefficients are
\begin{gather}\label{general_p_large_l_coeffs}
\begin{aligned}
& a_1^{f}=1\, ,\;a_2^{f}=\frac{i+1}{(1+i) p-i}-i\, , \;b_1^{f}=i\, , \;b_3^{f}=\frac{i-1}{(1+i) p-i}+1
\\ & a_1^{g}=1\, ,\;a_2^{g}=i+\frac{1+i}{(1+i) p - 1}\, , \;b_1^{g}=-i\, , \;b_3^{g}=1+\frac{1-i}{(1+i) p - 1}\;.
\end{aligned}
\end{gather}
The wave number, $k \in \mathbb{R}$ and $k>0$.\footnote{In the special case $p=\frac{1}{2}$ the discontinuous modes above, 
$\{\tilde{f}_k^{(p)}, \tilde{g}_k^{(p)} \}|_{p = \frac{1}{2}}$, 
are actually linear combinations of the modes that one obtains by performing the 
limiting procedure directly on the $\hat{f}^{(\frac{1}{2})}$ and $\hat{g}^{(\frac{1}{2})}$ modes in the pair of diamonds from Section \ref{disconthalf}.}


\subsection{Wightman function}

Denoting all  the modes collectively as $\tilde{u}_{i,k}=\left(\tilde{f}_k,\tilde{g}_k,\tilde{f}_k^{(p)},\tilde{g}_k^{(p)} \right)$, where $i=1,...,4$ labels the type of mode,  the field operator can be expanded as
\begin{equation}\label{field_operator_p_half}
\phi=\sum_{i=1}^4 \int_0^{\infty} dk \;(a_{i,k} \tilde{u}_{i,k} + a_{i,k}^{\dagger} \tilde{u}_{i,k}^*)\;,
\end{equation}
where $a_k^{\dagger}$ and $a_k$ are creation and annihilation operators respectively. 
The Wightman function is  
\begin{equation}\label{wight_p}
W_{p}(x,y)=\sum_{i=1}^4\int_{k_0}^{\infty} dk\; \tilde{u}_{i,k}(x) \tilde{u}_{i,k}^* (y)\,,
\end{equation}
where $k_0$ is an infrared cutoff,  needed because the theory is IR divergent,  as is the theory in Minkowski space.  In certain regions, this Wightman function  equals the  Minkowski Wightman function. Specifically, for all values of $p$, 
 $\left.W_{p}(x,y)\right|_{x,y\in R_i}=W_{\mathsf{Mink}}(x,y)$ for $i=1,3,5$ and $7$. The Wightman function differs from $W_{\mathsf{Mink}}$ when the arguments lie in regions spacelike to the singularity, or when $x$ and $y$ lie in different regions. It can also be shown that $W_{p}(x,y)=0$ if $x\in R_1$ and $y\in R_5$, 
 or $x\in R_3$ and $y\in R_7$: there is no correlation between the two disjoint pieces of the future/past of the singularity.

\subsection{Energy Density in the SJ State in the Infinite Trousers}\label{Stress-energy for SJ state in the infinite trousers}

The SJ Wightman function in the infinite trousers provides evidence that the energy density is zero everywhere except for the past and future lightcones of the singularity, for any $p$. 
Consider $x$ and $y$ in the same region, $R_i$, and not on the lightcone of $x_c$. Denote the UV cutoff Wightman function as $W^{\Lambda}_{p}(u,v;u',v')$, where $(u,v)$ and $(u',v')$ are the lightcone coordinates of  $x$ and $y$ respectively, and $\Lambda$ is a UV cutoff on the  $k$-integral in (\ref{wight_p}). Define the quantity
\begin{equation}
T^{\Lambda}_{p}(u,v;u',v'):=\frac{1}{2}(\partial_u \partial_{u'}+\partial_v \partial_{v'})W^{\Lambda}_{p}(u,v;u',v')\;.
\end{equation}
and the corresponding quantity $T^{\Lambda}_{\mathsf{Mink}}(u,v;u',v')$ for the Minkowski Wightman function, $W^{\Lambda}_{\mathsf{Mink}}(u,v;u',v')$. The expectation value of the energy density (on a surface of constant time) 
is then given by 
\begin{equation}
\big<0_{(p)}^\infty\big| T_{00}(x)\big|0_{(p)}^\infty\big>:=\lim_{\Lambda\rightarrow\infty}
\lim_{y\rightarrow x}\left(T^{\Lambda}_{p}(u,v;u',v')-T^{\Lambda}_{\mathsf{Mink}}(u,v;u',v')\right)\;,
\end{equation}
where $\big|0_{(p)}^\infty\big>$ is the SJ state in the infinite trousers.  
We already know that the difference is zero, before the limits are taken, in regions $R_i$, $i=1,3,5$ and $7$ because the SJ and Minkowski Wightman functions are equal there. 
It turns out that this difference is zero, before the limits are taken, in the other 
regions $R_i$, $i = 2,4, 6, 8$ as well. 

We can also see, at a formal level, that
there is a factor of $\delta(0)$  in the energy density on the lightcones from 
$x_c$.  Consider, without point splitting, 
\be\label{renorm_energy_density_infinite_p_half}
\big<0_{(p)}^\infty\big| T_{00} \big|0_{(p)}^{\infty} \big>_{reg}:=\big<0_{(p)}^\infty\big| T_{00} \big| 0_{(p)}^{\infty} \big> - \left< 0_{\mathsf{Mink}} \right| T^{\mathsf{Mink}}_{00} \left| 0_{\mathsf{Mink}} \right>\;,
\ee
where
\begin{equation}\label{energy_density_infinite_general_p}
\begin{aligned}
\big<0_{(p)}^\infty\big| T_{00} \big|0_{(p)}^\infty\big> & = \frac{1}{2} \int_0^{\Lambda} dk\, \Big( \partial_u \tilde{f}_k \partial_u \tilde{f}_k^*+\partial_v \tilde{f}_k \partial_v \tilde{f}_k^*
+ \partial_u \tilde{g}_k \partial_u \tilde{g}_k^*+\partial_v \tilde{g}_k \partial_v \tilde{g}_k^*  \\
& \left. +\partial_u \tilde{f}_k^{(p)} \partial_u \tilde{f}_k^{(p)*}+\partial_v \tilde{g}_k^{(p)} \partial_v \tilde{g}_k^{(p)*}
+ \partial_u \tilde{g}_k^{(p)} \partial_u \tilde{g}_k^{(p)*}+\partial_v \tilde{g}_k^{(p)} \partial_v \tilde{g}_k^{(p)*} \right)\;,
\end{aligned}
\end{equation}
and the Minkowski vacuum energy is
\begin{equation}\label{mink_vac_energy}
\begin{aligned}
\left< 0_M \right| T^M_{00} \left| 0_M \right> & = \frac{1}{2} \int_0^{\Lambda} dk\, \partial_u u_k \partial_u u_k^*+\partial_v v_k \partial_v v_k^* \\
& = \frac{1}{2}\int_0^{\Lambda} dk\, \frac{k}{2\pi}\;,
\end{aligned}
\end{equation}
where the Klein-Gordon normalised Minkowski space modes are $u_k:=\frac{1}{\sqrt{4\pi k}}e^{-iku}$ and $v_k:=\frac{1}{\sqrt{4\pi k}}e^{-ikv}$.

Let the point at which we evaluate this quantity have time coordinate less than zero.  
 In this region the modes $\tilde{f}_k^{(p)}$ and $\tilde{g}_k^{(p)}$ take the form
\begin{align}\label{kg_f_mode_p_thetas}
\tilde{f}_k^{(p)} & =\frac{1}{\sqrt{16\pi k}}\left(\left(-a_1^{f}\Theta(u)+a_2^{f}\Theta(-u)\right)e^{-iku}+\left(b_1^{f}\Theta(v)-b_3^{f}\Theta(-v)\right)e^{-ikv}\right)
\\
\label{kg_g_mode_p_thetas}
\tilde{g}_k^{(p)} & =\frac{1}{\sqrt{16\pi k}}\left(\left(-a_1^{g}\Theta(u)+a_2^{g}\Theta(-u)\right)e^{-iku}+\left(b_1^{g}\Theta(v)-b_3^{g}\Theta(-v)\right)e^{-ikv}\right)\;,
\end{align}
resulting in 
\begin{equation}\label{final_renorm_energy_density_infinite_general _p}
\begin{aligned}
\big<0_{(p)}^\infty\big| T_{00} \big|0_{(p)}^\infty \big>_{reg}=\frac{1}{2} \int_0^{\Lambda} & dk\,\left\lbrace \frac{k}{4\pi}+\frac{1}{8\pi k}\bigg[ k^2 \left(\Theta(u)^2+\Theta(-u)^2+\Theta(v)^2+\Theta(-v)^2\right) \right. \\
&+\frac{4}{1+2p(p-1)}\left(p^2\delta(u)^2+(1-p)^2\delta(v)^2 \right)\bigg] - \left.\frac{k}{2\pi}\right\rbrace\,.
\end{aligned}
\end{equation}
Integrating this over a segment of a constant time surface that does not intersect $u=0$ or $v=0$ gives $0$. However, if the 
surface intersects the $u=0$ ($v=0$) line then the result diverges 
unless $p=0$ ($p=1$).

For all $p$, the SJ state has divergent energy on both the past and future lightcones of the singularity. This is a consequence of the time reversal symmetry of the infinite trousers which is respected by the SJ state.

\section{Propagation and Nonunitarity}

Returning to the pair of diamonds, we can ask what ``propagation law''  the Green function corresponds to, in order to compare with previous work in~\cite{Copeland:1988tr}. We recall the usual evolution of initial data with a retarded Green function. Given a solution $f(x)$ of the field equation and its derivative on a spacelike hypersurface $\Sigma$ and a retarded Green function $G(x,y)$, the forward-propagated solution at a point $x$ in the future domain of dependence, $D^+(\Sigma)$, is 
\be
f(x) = \int_{\Sigma} d\Sigma_y^\mu\left[f(y)\nabla^y_\mu G(x,y)-G(x,y)\nabla^y_\mu f(y)\right]\,.\label{eq:forwardprop}
\ee

Consider now the pair of diamonds and retarded Green function $G_p(x,y)$. 
Take $\Sigma$ to be a spacelike surface that is a union of two disjoint pieces, $\Sigma=\Sigma_A\cup\Sigma_B$, where $\Sigma_A$ ($\Sigma_B$) goes from the left to right corners of diamond $A$ ($B$) and passes under the singularity: $\Sigma$ is as close as possible to a Cauchy surface. Using~\eqref{eq:forwardprop} we can propagate continuous initial data on $\Sigma$ to any point in its future. Given a solution on and to the past of $\Sigma$ we call the \emph{completely propagated solution} that which is generated by propagating to every point to the future of $\Sigma$ in this way.
For discontinuous initial data, the propagation law is not well defined, as it would involve derivatives of the discontinuous function multiplied by the discontinuous Green function.

If the initial data is continuous and even under the exchange $A\leftrightarrow B$, the
completely propagated solution is also even under the exchange. 
To see this, we first show that initial data corresponding to the $\hat{f}_k$ and $\hat{g}_k$ modes will propagate to the $\hat{f}_k$ and $\hat{g}_k$ modes respectively
everywhere. The result then follows because any solution that 
is even under the exchange is a linear combination of the $\hat{f}_k$ and $\hat{g}_k$ modes.

To see how initial data corresponding to an $\hat{f}_k$ or $\hat{g}_k$ mode propagates it suffices to consider the propagation of plane waves. Let us denote by $u^A_k(x)$ the function whose initial data is a right-moving plane wave on $\Sigma_A$ and which is zero on $\Sigma_B$, i.e. $u^A_k(y)=e^{-iku}\chi_{2,3,4}(x)$. (\ref{eq:forwardprop})  evolves $u^A_k$ to $+p$ for $x\in R_1$ and to $e^{-iku}-p$ for $x\in R_5$. 

We can also specify the initial data on $\Sigma$ for the following plane waves: $u^B_k(x)=e^{-iku}\chi_{6,7,8}(x)$, $v^A_k(x)=e^{-ikv}\chi_{2,3,4}(x)$ and $v^B_k(x)=e^{-ikv}\chi_{6,7,8}(x)$. $u^B_k(x)$ and $v^B_k(x)$ are zero on $\Sigma_A$, and $v^A_k(x)$ is zero on $\Sigma_B$. Their corresponding completely propagated solutions are:
\bea
\label{eq:u1u2}
u^A_k(x)&=e^{-iku}\chi_{2,3,4,5}(x) + p\left[\chi_5(x)-\chi_1(x)\right]\\
u^B_k(x)&=e^{-iku}\chi_{1,6,7,8}(x) + p\left[\chi_1(x)-\chi_5(x)\right]\;,
\eea
and
\bea\label{eq:v1v2}
v^A_k(x)&=e^{-ikv}\chi_{1,2,3,4}(x) + (1-p)\left[\chi_5(x)-\chi_1(x)\right]\\
v^B_k(x)&=e^{-ikv}\chi_{5,6,7,8}(x) + (1-p)\left[\chi_1(x)-\chi_5(x)\right].
\eea
Taking linear combinations of the above modes, one can verify that the $\hat{f}_k$ and $\hat{g}_k$ modes ``propagate into themselves" in the sense described above.

To compare this to the results in~\cite{Copeland:1988tr} we recall how the pair of diamonds was cut out from the trousers. The modes on the pair of diamonds corresponding to the natural ``right-moving plane waves in the trunk'' from~\cite{Copeland:1988tr} with periodic boundary conditions take the form $u_k^A(x)+(-1)^n u_k^B(x)$ with $k=\sqrt{2}n\pi/\lambda$ in our conventions (the factor of $\sqrt2$ here arises from our definition of the light-cone coordinates). For even $n$, the constant terms in~\eqref{eq:u1u2} cancel. For odd $n$, they add up, leading to opposite constant terms $\pm 2p$ in the causal futures of the singularity in the left/right legs. Similar statements apply to left-moving incoming modes. This corresponds precisely to the one-parameter family of propagation laws found in~\cite{Copeland:1988tr}, which the authors arrived at by demanding the conservation of what they call the ``Klein-Gordon inner product'' under the evolution past the singularity. Our parameter $p$ is related to the parameter $A$ in~\cite{Copeland:1988tr} via $p=\frac12(1+A)$.

\begin{figure}[t!]
\centering
\includegraphics[
clip=true,
width=\textwidth]
{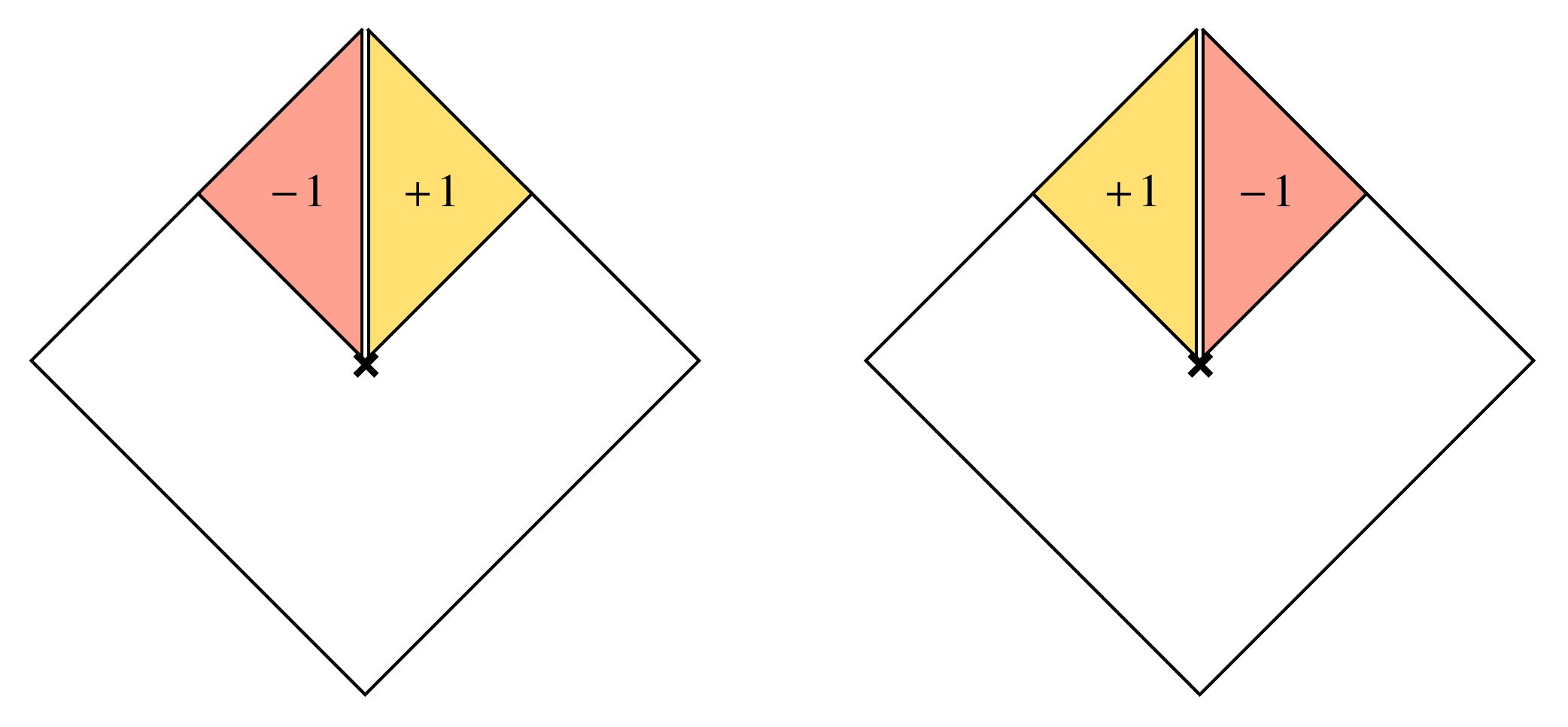}
\caption{Illustration of the $\Gamma_0(x)$ function. The function is zero in the white regions.}
\label{fig:gamma0}
\end{figure}

\begin{figure}[t!]
\centering
\includegraphics[
clip=true,
width=\textwidth]
{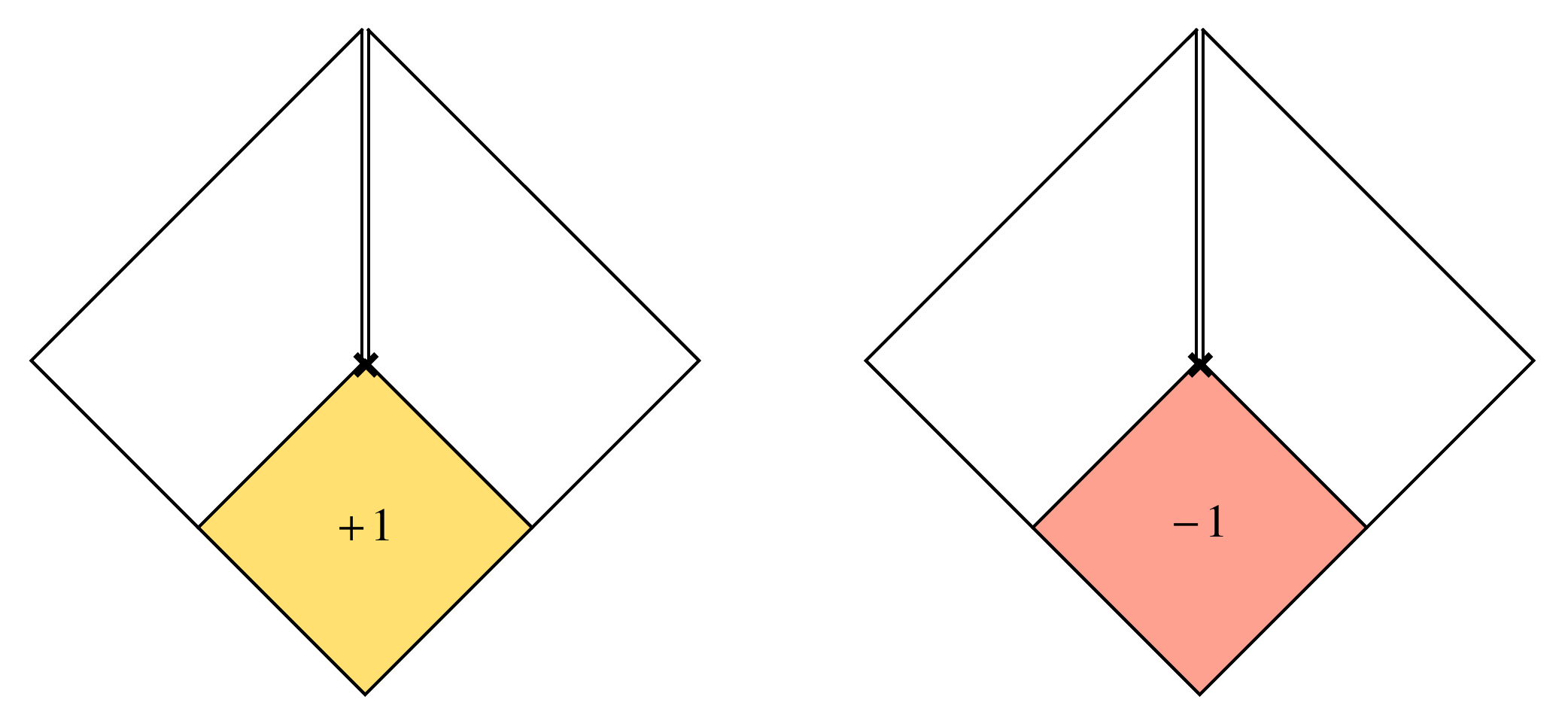}
\caption{Illustration of the $\Gamma(x)$ function. The function is zero in the white regions.}
\label{fig:gamma}
\end{figure}

At the end of~\cite{Copeland:1988tr} the authors mention certain discontinuous functions, which they call $\gamma_0(x)$ and $\gamma(x)$, that violate the propagation rule, and ask whether they are required to form a complete set of modes. The analogous functions in $\mathcal{M}$ are $\Gamma_0(x)=\chi_1(x)-\chi_5(x)$ and $\Gamma(x)=\chi_3(x)-\chi_7(x)$ as illustrated in Figure \ref{fig:gamma0} and \ref{fig:gamma} respectively. Each function satisfies the requirements for a solution, and so is expressible as a linear combination of the SJ modes and this means that in the pair of diamonds the notion of ``propagation'' becomes ill-defined. Solutions $f(x)$  and $f(x) +  
\lambda \Gamma_0(x)$, where $\lambda$ is a constant, share the same initial data. 
Similarly, $f(x)$ and $f(x) + \lambda \Gamma(x)$ have the same final data. 

\subsection{ Nonunitarity}

The  ambiguity in the notion of propagation 
 indicates that the theory in the pair of diamonds is nonunitary. We will see that this can be expressed as the algebra of 
 observables, $\mathfrak{A}_-$, associated to the past null boundary, $N_-$, being a 
 strict subset of the algebra of observables, $\mathfrak{A}$, for the full spacetime. 
  
\begin{figure}[t]
\centering
\includegraphics[
clip=true,
width=\textwidth]
{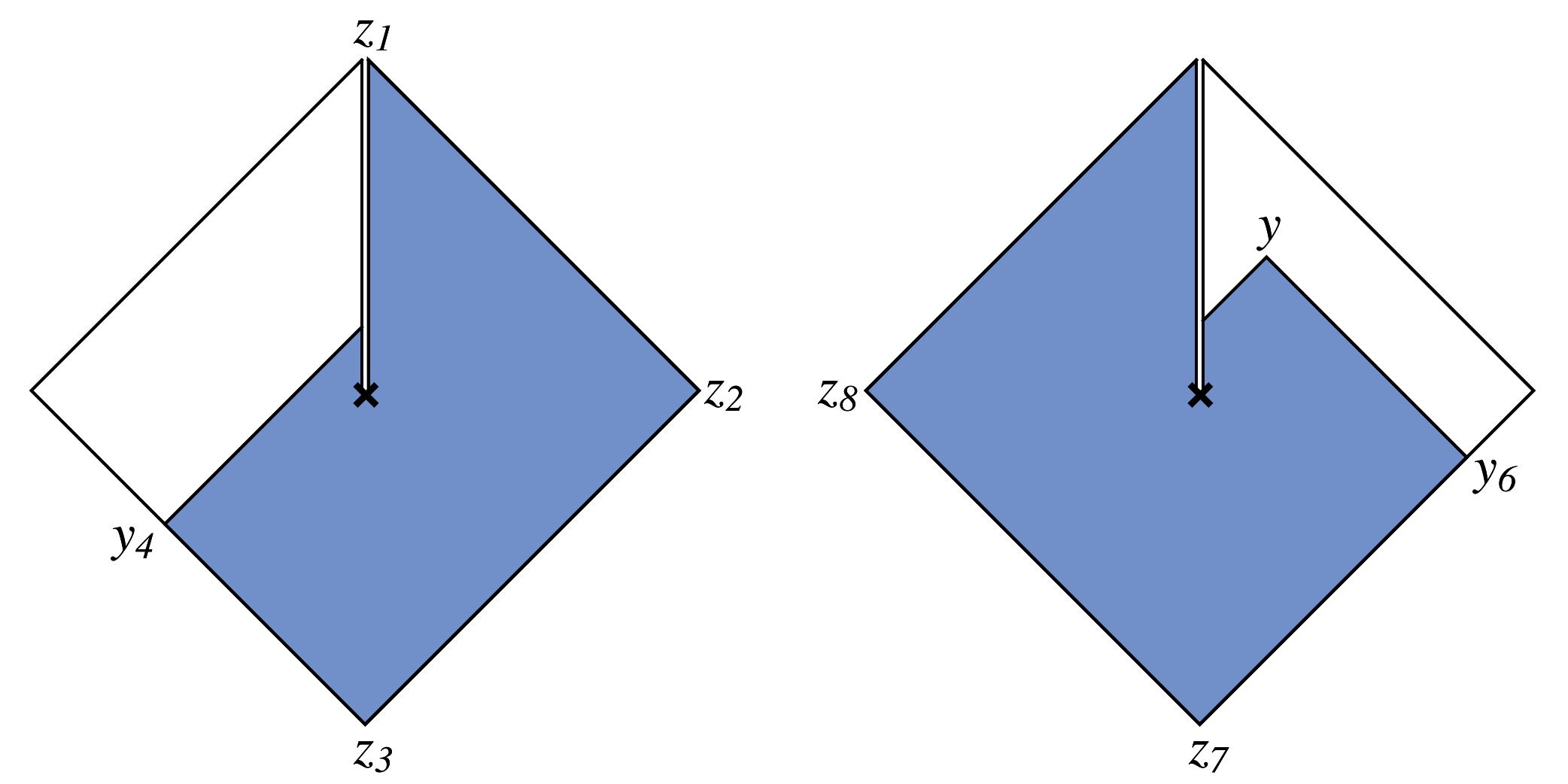}
\caption{The double diamond for equation \eqref{double_diamond_2}.}
\label{fig:unitarity-region-2}
\end{figure}

 Let the vertices of the pair of diamonds be labelled $z_1, z_2, \dots z_8$ in clockwise order starting from $z_1$ which is the top vertex of region $R_1$, as shown in 
 Figure \ref{fig:unitarity-region-2}. $z_i \in R_i$ for all $i$. Given any point $x$ not in the causal future of the singularity, $x_c$, the equation of motion (\ref{diamondeom}) for a diamond with $x$ at its top vertex and the other three vertices on the past null boundary $N_-$ shows that $\phi(x)$ is determined by values of $\phi$ on $N_-$.
\begin{figure}[t]
\centering
\includegraphics[
clip=true,
width=\textwidth]
{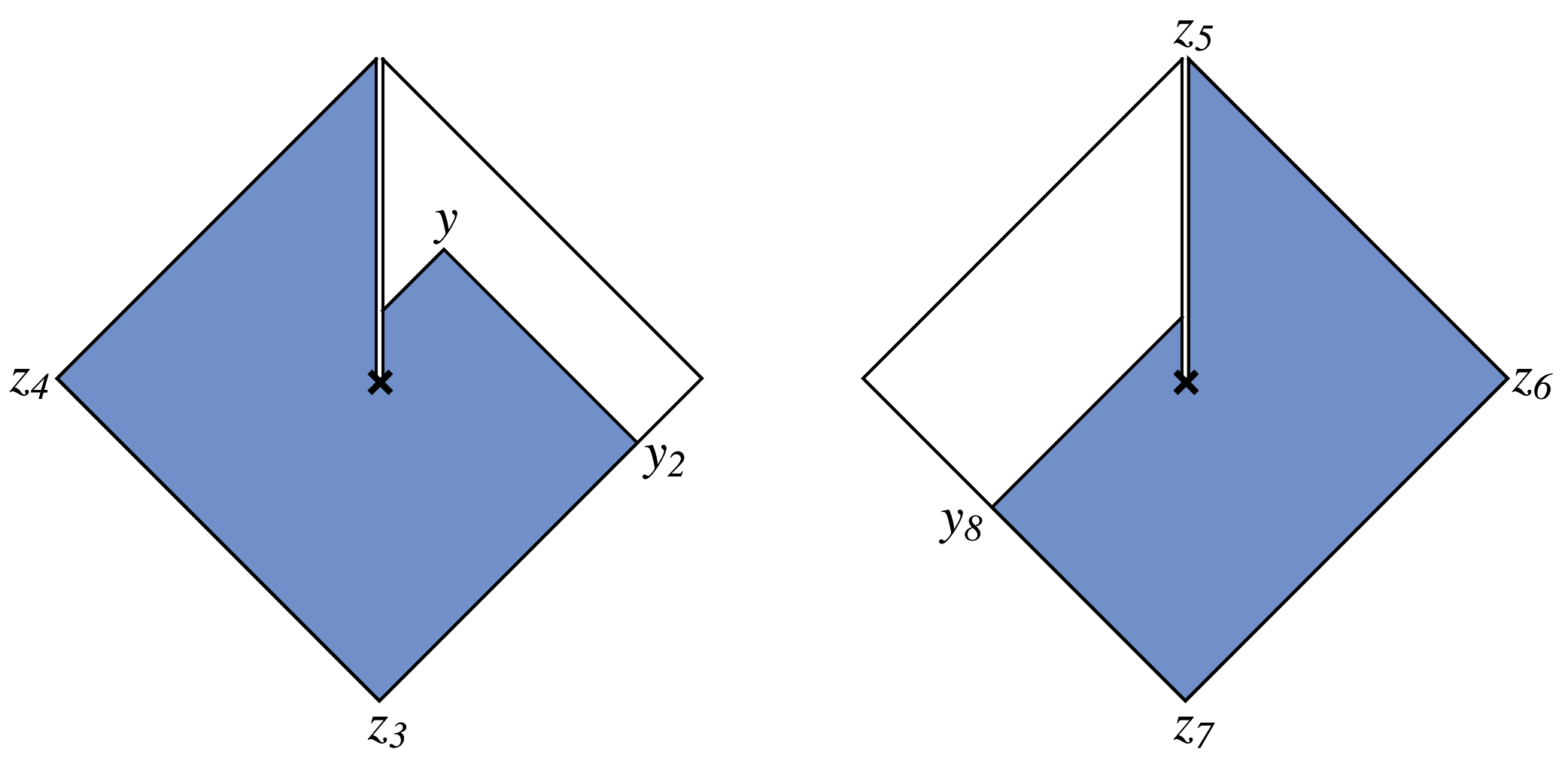}
\caption{The double diamond for equation \eqref{double_diamond_1}.}
\label{fig:unitarity-region-1}
\end{figure}
However, if $y\in R_1$ then $\phi(y)$ is not specified by the initial data on $N_-$ since, using equation of motion (\ref{ddiamondeom}) for the double diamond shown in Figure \ref{fig:unitarity-region-1}, 
\begin{equation}\label{double_diamond_1}
\phi(y) = \phi(y_2) - \phi(z_3) + \phi(z_4) - \phi(z_5) + \phi(z_6) - \phi(z_7) + \phi(y_8)\;,
\end{equation}
where $y_2 \in R_2\cap N_-$ and $y_8 \in R_8\cap N_-$ are the points shown in Figure \ref{fig:unitarity-region-1} and $z_5 \notin N_-$. 

Similarly, if $y \in R_5$, then the double diamond in Figure \ref{fig:unitarity-region-2} gives
\begin{equation}\label{double_diamond_2}
\phi(y) = \phi(y_6) - \phi(z_7) + \phi(z_8) - \phi(z_1) + \phi(z_2) - \phi(z_3) + \phi(y_4)\;,
\end{equation}
where $y_4 \in R_4\cap N_-$ and $y_6 \in R_6\cap N_-$ are the points shown in Figure \ref{fig:unitarity-region-2} and $z_1 \notin N_-$. 
In both cases $\phi(y)$ is not specified by data on $N_-$. However, the extra data needed is not $\phi(z_1)$ and $\phi(z_5)$ since, the equation of motion from the double diamond that is the whole pair of diamonds implies their sum is specified by data on $N_-$:
\begin{equation}\label{phi_sum}
\phi(z_1) + \phi(z_5)  =  \phi(z_2) - \phi(z_3) + \phi(z_4) + \phi(z_6)  - \phi(z_7) + \phi(z_8)\,.
\end{equation}
 Therefore, only $\Phi_+ := \phi(z_1) - \phi(z_5)$ is needed to complement $\phi$ on $N_-$.
 
 Similarly, a solution $\phi$  is specified by data on the future null boundary, $N_+$, 
 together with $\Phi_- := \phi(z_3) - \phi(z_7)$. 
 
Thus, $\Phi_+$ ($\Phi_-$) and all operators generated from it are missing from the algebra $\mathfrak{A}_-$ ($\mathfrak{A}_+$). The structural relationship between $\mathfrak{A}_-$, $\mathfrak{A}_+$ and $\mathfrak{A}$ remains to be worked out. Here we just note that 
\begin{gather}\label{phi_commutator}
\begin{aligned}
[\Phi_+,\Phi_-] & =[\phi(z_1),\phi(z_3)]-[\phi(z_5),\phi(z_3)]-[\phi(z_1),\phi(z_7)]+[\phi(z_5),\phi(z_7)]
\\
& = i( \Delta(z_1,z_3)-\Delta(z_5,z_3)-\Delta(z_1,z_7)
+\Delta(z_5,z_7) )
\\
& = i (1-2p)\;.
\end{aligned}
\end{gather}
so the operators commute for $p=\frac{1}{2}$. 

\section{Outlook}

Trying to make sense of quantum field theory on a topology changing background not 
only advances the study of topology change but requires us to think afresh about QFT 
and its foundations. As the SJ formalism for free quantum field theory depends only on spacetime causal order and the retarded Green function, it is straightforward, at least in principle, to apply it to the pair of diamonds, a topology changing spacetime. The surprise was that the SJ modes could be found, and the Wightman function constructed, explicitly. Some of these modes are discontinuous across the future and past lightcones of the singularity and this discontinuity gives rise to a divergence in the energy density on these null lines, 
confirming the expectation arising from past work by Anderson and DeWitt and 
by Copeland et al. A similar conclusion was reached by examining the limiting case of the infinite trousers. As the SJ state is time reversal symmetric, the divergences appear on both the past and future lightcones of the singularity in contrast to 
previous work.  We have also found a relation between the SJ framework based on the Green function and previous work by Copeland et al. by analysing the concept of propagation forward in time. In a unitary theory,
if spacetime region $X$ is in the domain of dependence of region $Y$ then 
then the corresponding algebras of observables are related
by $\mathfrak{A}_X \subseteq \mathfrak{A}_Y$. However we have seen that this fails in the pair of diamonds: the future boundary, $N_+$, is in the domain of dependence of the past boundary, $N_-$, but the corresponding algebra $\mathfrak{A}_+$  contains an operator that is not in $\mathfrak{A}_-$

How should these results be viewed by those who  believe that topology change should
be part of full quantum gravity? One could argue that since topology change is 
expected to be a quantum gravity effect we should study it in the context of a background spacetime with no structure at the Planck scale, 
for example a causal set and this would be interesting to do.
It is possible, though, that these results and the previous work are 
telling us  
that topology-change of the trousers type is disallowed whilst leaving the question of other types of topology-change very much open. The transition in the trousers belongs to the class of topology-changes in which the spacetime exhibits ``causal discontinuity''~\cite{Dowker:1997hj,Dowker:1999wu} where the causal past or future of a point  changes discontinuously as the point moves across the past or future lightcones of the singularity. The authors of~\cite{Louko:1995jw} found evidence that causally discontinuous topology changing processes in $1+1$ dimensions are suppressed in a sum-over-histories, while causally continuous ones are enhanced. Such observations lend support to Sorkin's conjecture that the pathology of infinite energy production occurs in a topology-changing spacetime if and only if it is causally discontinuous. 

It would be very interesting therefore to study the type of topology change in 3+1 dimensions with a singularity with Morse signature $(++--)$ which is  causally continuous. This type of topology change is particularly interesting in 3+1 dimensions because, given any two closed connected 3-manifolds, there exists a cobordism between them which  admits a Lorentzian metric with only these types of singularity. It would be interesting to study the SJ theory of a scalar field in such a spacetime. If it can be shown that the SJ Wightman function is well behaved in a case like this, it would be strong evidence that the pathology of divergent energy production is associated only
with the trousers. 

\section*{Acknowledgements} This work is supported by STFC grant ST/L00044X/1. Research at Perimeter Institute is supported by the Government of Canada through Industry Canada and by the Province of Ontario through the Ministry of Research and Innovation. MB acknowledges support from NSF Grant No. CNS-1442999. IJ is supported by the EPSRC. MB and IJ thank Perimeter Institute for hospitality while this work was being completed. The authors would also like to thank Henry Wilkes for his helpful suggestions towards this work.
\begin{appendices}
\section{Zero Eigenvalue Eigenfunctions Are Not Solutions}\label{Zero Eigenvalue Eigenfunctions Are Not Solutions}

We first derive a simple formula that must be satisfied by a zero eigenvalue eigenfunction (ZEE).

Consider $x$ in diamond $A$ with coordinates $(L,v')$ with $v'<0$. For a function $f$, $i\Delta_pf=0$ implies $\int_{-L}^{v'}dv\int_{-L}^Ldu f(u,v)=0$. Differentiating this expression with respect to $v'$ implies  $\int_{-L}^Ldu f(u,v')=0$, for all $v'<0$. Similarly, all integrals of $f$ along lines of constant $u$ vanish. So, 
\begin{equation}\label{integral_null_line}
\int_{-L}^L du f(u,v')=0\;\;\;\text{and}\;\;\int_{-L}^L dv f(u',v)=0\;\;\forall\;u',v'\neq 0\,.
\end{equation}

We say a nonzero function $f$ is a ZEE if it satisfies~\eqref{integral_null_line}. We say an element of ${L}^2(M)$ is a ZEE if it has a nonzero representative which satisfies~\eqref{integral_null_line}.

\noindent{\textbf{Claim:}}
If  $[f]\in {L}^2(M)$ is both a ZEE and a solution, then it has a representative function that is both a ZEE and a solution.

\noindent{\textbf{Proof:}} 
It suffices to show that a representative function of $[f]$ which is a solution, and which  we might as well call $f$, can be changed on a set of measure zero,  so that it satisfies~\eqref{integral_null_line}, whilst remaining a solution. Recall the conditions for a function to be a solution are $ \mathfrak C^{D} f= 0$ for every diamond, $D$, that doesn't contain $x_c$ and $ \mathfrak C^{DD} f= 0$ for every double diamond, $DD$, that contains $x_c$. 

The function $f(x)$ can only fail~\eqref{integral_null_line} on a set of lines of measure zero, since $[f]$ is a ZEE. On one such null line the integral of $f(x)$ will be some non-zero real number, $\eta$. We can alter $f(x)$ by subtracting from it the function that is $\frac{\eta}{2L}$ along that null line and zero everywhere else. The resulting function, $\tilde{f}(x)$, now satisfies~\eqref{integral_null_line} on that particular null line and $\tilde{f}(x)$ still satisfies the conditions for it to be a solution. We can continue to 
adjust the function in this way for all of the null lines on which $f(x)$ failed~\eqref{integral_null_line}. The resulting function will be both a solution and a ZEE.

\noindent{\textbf{Claim:}}  $[h]\in {L}^2(M)$ cannot be both a ZEE and a solution.

\noindent{\textbf{Proof:}} Let $[h]\in {L}^2(M)$ be both a ZEE and a solution
and let the representative, $h$, be both a ZEE and a solution. We will prove that $h \sim 0$, the 
zero function. 

$h(x)$ satisfies $\mathfrak{C}^{D} h=0$ for all diamonds $D$ that do not contain $x_c$. Take such a $D$ in diamond $A$ with corners $x_1,x_2,x_3$ and $x_4$ that have light-cone coordinates $(u,v),(-L,v),(-L,-L)$ and $(u,-L)$ respectively, where $u\in [-L,L]$ and $v\in [-L,0)$. With these coordinates the equation $\mathfrak{C}^{D} h=0$ becomes
\begin{equation}\label{simple_eom_with_coords}
h(u,v)-h(-L,v)+h(-L,-L)-h(u,-L)=0\,.
\end{equation}
Integrating this along $u$, and using $\int_{-L}^L du\, h(u,v)=0$ for all $v\neq 0$  
gives   
 $h(-L,v)=h(-L,-L)$. This is true for all $v<0$ and so $h(x)$ is
constant along the line of $u=-L$ for $v<0$. The same reasoning shows that on the line of $v=-L$ for $u<0$ the function must equal the same constant, which we call $C$.

Using this in~\eqref{simple_eom_with_coords} implies that $h(u,v)=C$ if $u,v<0$ in diamond $A$, \textit{i.e.} in region $R_3$.  Similar reasoning shows that $h(x)$ is constant in the interior of each region $R_i$ for $i=1,...,8$. Given that $h(x)$ is a ZEE it must satisfy ~\eqref{integral_null_line} which implies the constants must be equal in magnitude in each region with alternating signs as one traverses the regions $R_1$ to $R_8$ in order. Therefore $h(x)=C\sum_{i=1}^8 (-1)^{i-1}\chi_i(x)$. The equation of motion, $\mathfrak{C}^{DD} h=0$, then implies that $C=0$.

\section{Sum of Squares of Eigenvalues}\label{appendix_sum_of_square_evals}

\subsection{ $p\neq \frac{1}{2}$}
 The sum over all the positive and negative eigenvalues is
\begin{equation}\label{all_evals_sum}
\sum_{\text{all modes}}\lambda_k^2=\sum_{\text{cont.}}\lambda_k^2 +\sum_{\text{discont.}}\lambda_k^2\;,
\end{equation}
where the first sum on the right is over the continuous copy modes  ($\hat{f}_k$, $\hat{g}_k$ and their complex conjugates), and the second sum is over the discontinuous modes ($\hat{f}_k^{(p)}$, $\hat{g}_k^{(p)}$ and their complex conjugates). The sum over the continuous modes equals $2L^4$ \cite{Johnston:2010su}. As the eigenvalues come in positive and negative pairs  the second sum equals twice the sum over just the positive eigenvalues. Then
\begin{equation}\label{all_evals_sum_partial_evaluation}
\sum_{\text{all modes}}\lambda_k^2=2L^4 +2\sum_{\substack{\text{pos.}\\ \text{discont.}}}\lambda_k^2=2L^4 +2 {L}^2\left(\sum_{\substack{\text{pos.}\\ +}}k^{-2}+\sum_{\substack{\text{pos.}\\ -}}k^{-2}\right)\;,
\end{equation}
where the last expression uses $\lambda_k=\frac{L}{k}$, and the sum
 over the positive eigenvalues of the discontinuous modes is split into two sums over $k>0$ satisfying~\eqref{general_p_eigenval_eqns} with the ``$+$" and ``$-$" signs respectively. 


The ``$+$" sign in~\eqref{general_p_eigenval_eqns} gives the following transcendental equation for $k$:
\begin{equation}\label{example_transcendental_eqn}
\left((kL)^2+2\right) \cos (kL)+kL (2+ kL (1-2 p)) \sin (kL)-2=0\;.
\end{equation}
For $p\neq \frac{1}{2}$, equation~\eqref{example_transcendental_eqn} has both positive and negative roots with no degeneracy.
One can verify that the set of negative roots of~\eqref{example_transcendental_eqn} is equal to the set of positive roots of~\eqref{general_p_eigenval_eqns} with the ``$-$" sign chosen. This means that the last two sums in~\eqref{all_evals_sum_partial_evaluation} can be written as a single sum over all roots (positive and negative) of~\eqref{example_transcendental_eqn}, which we write as $\sum_i{k_i}^{-2}$.

Taylor expanding $\cos(kL)$ and $\sin(kL)$ about $k=0$  in~\eqref{example_transcendental_eqn} gives
 \begin{equation}\label{eval_eqn_polynomial}
2 (kL)^2 +(1-2p)(kL)^3-\frac{3}{4}(kL)^4+\mathcal{O}(k^5)=0\,.
\end{equation}
We can think of~\eqref{eval_eqn_polynomial} as an infinite degree polynomial, if we imagine continuing the expansion forever. We want to evaluate a sum over a particular power of the roots of this infinite polynomial. To do this we require a result from finite degree polynomials.

Expressing a polynomial of finite degree in terms of its roots,
\begin{equation}\label{polynomial}
\alpha_n x^n + \alpha_{n-1} x^{n-1} +...+\alpha_1 x + \alpha_0 = \alpha_n (x-x_1)(x-x_2)...(x-x_n)\,,
\end{equation}
one can verify Vieta's formulae. From these it is straightforward to show that
\begin{equation}\label{coeffs_sum_of_reciprocal_squared_roots}\
\left(-\frac{\alpha_1}{\alpha_0}\right)^2-2\frac{\alpha_2}{\alpha_0} =\frac{1}{{x_1}^2}+\frac{1}{{x_2}^2}+...+\frac{1}{{x_n}^2}\;.
\end{equation}
Such formulae are extended in~\cite{speigel1953summation} to 
infinite polynomials such as ~\eqref{example_transcendental_eqn}. Dividing~\eqref{eval_eqn_polynomial} by $(kL)^2$  gives 
\begin{equation}\label{x_polynomial}
2+(1-2p)kL-\frac{3}{4}(kL)^2+\mathcal{O}(k^3)=0\;,
\end{equation}
so that $\alpha_0=2$, $\alpha_1=(1-2p)L$, and $\alpha_2=-\frac{3{L}^2}{4}$, which gives $\sum_i{k_i}^{-2}={L}^2(1-p(1-p))$. The sum of the squares of all the positive and negative eigenvalues of $i\Delta_p$ is then
\begin{equation}\label{sum_of_square_of_evals_p_final}
\sum_{\text{all modes}}\lambda_k^2=2L^4 +2L^4\left(1-p(1-p) \right)=2L^4\left(2-p(1-p) \right)\;.
\end{equation}
This agrees with~\eqref{eq:hsnormpod}, which means that we have all the eigenfunctions of $i\Delta_p$.

\subsection{$p=\frac{1}{2}$}
The sum  over the eigenvalues is again split into sums over the continuous and discontinuous modes. The sum over the continuous modes gives $2L^4$, as
before, and the sum over the discontinuous modes can be written as twice the sum over the positive eigenvalues. Then, 
\begin{equation}\label{p_half_all_evals_sum_partial_evaluation}
\sum_{\text{all modes}}\lambda_k^2=2L^4 +2\sum_{\substack{\text{pos.}\\ \text{discont.}}}\lambda_k^2=2L^4 +2 {L}^2\left(\sum_{\substack{\text{pos.}\\ f}}k^{-2}+\sum_{\substack{\text{pos.}\\ g}}k^{-2}\right)\;,
\end{equation}
where, in the last two sums, we have used $\lambda_k=\frac{L}{k}$ with $k>0$ satisfying~\eqref{phalfevaleqn}. The sum over the discontinuous modes is split into two sums over the eigenvalues of the $\hat{f}_k^{(\frac{1}{2})}$ and $\hat{g}_k^{(\frac{1}{2})}$ modes respectively.


Since the transcendental equation~\eqref{phalfevaleqn} for $k$ is the same for the two sets of modes $\hat{f}_k^{(\frac{1}{2})}$ and $\hat{g}_k^{(\frac{1}{2})}$, the last two sums in~\eqref{p_half_all_evals_sum_partial_evaluation} are equal. The transcendental equation  is
\be\label{phalfevaleqn_appendix}
(2 + (kL)^2) \cos (kL) + 2 kL \sin (kL) - 2 = 0\,.
\ee
The roots of this equation come in positive/negative pairs of the same absolute value, and so the sum over the positive roots will be equal to half the sum over all the roots. Hence the last term in brackets in~\eqref{p_half_all_evals_sum_partial_evaluation} is equal to a sum over all the roots of~\eqref{phalfevaleqn_appendix}, which we write as $\sum_i{k_i}^{-2}$.

The Taylor expansion of~\eqref{phalfevaleqn_appendix} around $k=0$ is
\be\label{eval_eqn_p_half_expansion}
2(kL)^2-\frac{3}{4}(kL)^4+\mathcal{O}(k^6)= 0\;.
\ee
Dividing by $(kL)^2$ we find $\alpha_0=2$, $\alpha_1=0$ and $\alpha_1=-\frac{3{L}^2}{4}$ and hence $\sum_i{k_i}^{-2}=\frac{3{L}^2}{4}$. The sum over all the eigenvalues is
then 
\begin{equation}\label{p_half_all_evals_sum_final}
\sum_{\text{all modes}}\lambda_k^2=2L^4 +2{L}^2\left(\sum_{\substack{\text{pos.}\\ f}}k^{-2}+\sum_{\substack{\text{pos.}\\ g}}k^{-2}\right)=2L^4+2{L}^2\sum_i{k_i}^{-2}=\frac{7L^4}{2}\;,
\end{equation}
which is equal to the right hand side of~\eqref{eq:hsnormpod} with $p=\frac{1}{2}$.

\end{appendices}

\newpage
\bibliography{biblio}
\bibliographystyle{jhep}

\end{document}